# Water Structure and Electric Fields at the Interface of Oil Droplets

Lixue Shi[1,2†], R. Allen LaCour[3,4†], Xiaoqi Lang[1], Joseph P. Heindel[3,4], Teresa Head-Gordon[3,4*], Wei Min[1*]

[1]Department of Chemistry, Columbia University, New York, NY, 10027

[2]Shanghai Xuhui Central Hospital, Zhongshan-Xuhui Hospital, and Shanghai Key Laboratory of Medical Epigenetics, International Co-laboratory of Medical Epigenetics and Metabolism, Institutes of Biomedical Sciences, Shanghai Medical College, Fudan University, Shanghai 200032

[3]Kenneth S. Pitzer Theory Center and Department of Chemistry, University of California, Berkeley, CA, 94720 USA

[4]Chemical Sciences Division, Lawrence Berkeley National Laboratory, Berkeley, CA, 94720 USA

[†]Authors contributed equally

*E-mail: wm2256@columbia.edu and thg@berkeley.edu

## Abstract

Mesoscale water-hydrophobic interfaces are of fundamental importance in multiple disciplines, but their molecular properties have remained elusive for decades due to experimental complications and alternate theoretical explanations. Surface-specific spectroscopies, such as vibrational sum-frequency techniques, suffer from either sample preparation issues or the need for complex spectral corrections. Here, we report on a robust "in solution" interface-selective Raman spectroscopy approach using multivariate curve resolution to probe hexadecane-in-water emulsions. Computationally, we use the recently developed monomer-field model for Raman spectroscopy to help interpret the interfacial spectra. Unlike with vibrational sum-frequency techniques, our interfacial spectra are readily comparable to the spectra of bulk water, yielding new insights. The combination of experiment and theory show that the interface leads to reduced tetrahedral order and weaker hydrogen bonding, giving rise to a substantial water population with dangling OH at the interface. Additionally, the stretching mode of these free OH experiences a ~80 $cm^{-1}$ red-shift due to a strong electric field which we attribute to the negative zeta potential that is general to oil droplets. These findings are either opposite to, or absent in, the molecular hydrophobic interface formed by small solutes. Together, water structural disorder and enhanced electrostatics are an emergent feature at the mesoscale interface of oil-water emulsions, with an estimated interfacial electric field of ~35-70 MV/cm that is important for chemical reactivity.

## Introduction

While water is infamous as an anomalous liquid, interfacial water also displays fundamentally rich behavior[1-4], which has important implications in chemistry, biology, geology, and engineering. However, many properties of water at hydrophobic interfaces, such as the water/air or water/oil interface, remains hotly contested[5-11]. Contradictory conclusions about the most basic properties of interfacial water, such as whether the hydrogen bonded network of water is more or less ordered at a hydrophobic interface,[5-9] restrict our understanding of absorption, binding, and reactivity at such interfaces. This controversy stems from the difficulty in measuring interfacial molecular properties with even the most advanced experimental approaches.

Vibrational sum frequency generation (VSFG) has been the primary spectroscopic technique for this task, as its second-order nonlinear optical response gives it exquisite surface selectivity. Indeed, VSFG in the reflective geometry was employed to investigate planar oil interfaces, including water/alkane and water/$CCl_4$, as early as 1994[5,7,8]. However, contradictory conclusions have been drawn about the structure of water at hydrophobic interfaces[5-9] (Supplementary Note 1). These issues likely stem in part from poorly controlled sample-preparation issues such as partial wetting (resulting in formation of islands or lenses) and the presence of surface-active impurities (trace amount of impurities cover the planar interface)[6,7]. Furthermore, recent work indicates that some VSFG contributions may result from bulk-like water located close to the interface[12-14], implying that it may not be entirely interface specific.

To obtain a superior sample, researchers replaced planar interfaces with oil droplets dispersed in water, i.e. metastable but long-lived emulsions[15-17]. The surface to volume ratio is increased by over 1000 times, which drastically reduces the effect of any unwanted impurities, and better wetting is achieved due to an ultrasonication process. Yet the absence of a macroscopic interface in the emulsion prohibits the use of VSFG. To this end, vibrational sum frequency scattering (VSFS) with a forward-scattering geometry has been employed[11,18,19]. However, the infrared radiation needed for vibrational excitation is inevitably attenuated when going through the aqueous medium, affecting the resulting VSFS spectrum in a frequency-dependent manner[20,21]. Although researchers have attempted to compensate for this absorptive effect, the employed multi-parameter models are so complicated that opposite conclusions have been reached even when independent groups employ essentially the same technique (Supplementary Note 1)[20,21].

Parallel to these issues are fundamental questions regarding oil droplets, which are widely present in nature and a model system for hydrophobicity. Oil droplets in water are known to have a large negative zeta potential[15,22-24] which implies that they carry charge. As is the case for other electrostatically stabilized colloids, a finite zeta potential is necessary for the emulsion's long-term stability[15]. The origin of the negative zeta potential is contested, however, with various explanations including interfacial hydroxide adsorption and charge transfer between the water and the oil[11,25-27]. While there is a lack of a consensus for its molecular origin, the effect of charge found in oil droplets is a feature that distinguishes it from oil-water planar interfaces.

Due to the aforementioned difficulties with VSFG and VSFS, here we develop a new method for probing the interface in an oil-in-water emulsion system. Our inspiration comes from hydration-shell Raman spectroscopy, which employs high-resolution Raman spectroscopy in tandem with multivariate curve resolution (Raman-MCR)[28-30]. Using Raman-MCR, one can separate a measured Raman spectra into a pure solvent spectrum and a solute-correlated (SC) spectrum which encompasses solute vibrations in addition to solvent molecules whose vibrational modes are perturbed by the solute[30]. The resulting SC spectrum provides robust "in solution" interface-selectivity. While Raman-MCR was developed for small molecules[28-32], here we report an advance for extracting the interfacial spectrum on the mesoscopic scale, for the first time.

Unlike with VSFG and VSFS, our interfacial spectra are readily comparable to the bulk solution, yielding new insights. Compared to bulk water, the characteristic shoulder peak of OH stretching at 3250 $cm^{-1}$, which is associated with strong hydrogen bonding and structural order, is reduced to a vanishing intensity at the oil-water mesoscale interface. This is in stark contrast to the observation on small aqueous hydrophobic solutes[29,33-35]. Using molecular dynamics simulations and the newly developed monomer-field model for Raman spectroscopy[36], we also find a more disordered structure with weaker hydrogen bonds for the water at the oil-water interface compared to bulk water, which also reduces the shoulder at 3250 $cm^{-1}$ in our modelling of the SC spectrum.

The mesoscopic interface also gives rise to a substantial water population (~25%) with dangling (i.e., free) OH, and partly a consequence of a disordered and weakened hydrogen bonding network. Importantly, its stretching frequency (~3575 $cm^{-1}$) in the experimental SC spectrum appears to have experienced a ~80 $cm^{-1}$ red-shift from the expected position around 3650 $cm^{-1}$. We estimate from experiment and theory that a strong electric field strength of 35-70 MV/cm shall exist to red-shift the free OH to quantitatively agree with the observed sub-peak, which we attribute to the negative zeta potential of the oil droplet interface. Thus, the mesoscopic interface and the local charge of oil droplets *together* dictate the interfacial electric field properties[37,38], which has fundamental and broad implications for many chemical systems, and are emerging as the key factors contributing to substantial rate accelerations documented in microdroplet chemistry[39,40].

## Results

### Raman-MCR spectroscopy of oil droplets in water

Raman spectroscopy is appealing for hydration-shell spectroscopy because 1) the Raman cross-sections of liquid water are insensitive to changes in H-bond strength and 2) the timescale of Raman transitions is faster than the exchange of molecules between the solvation-shell and bulk solvent. Furthermore, it does not require prior knowledge or assumptions regarding the shape of the spectroscopic signals, and thus it has proven a powerful strategy for studying hydration shell of small molecules[28-32]. Within the Raman-MCR formalism, the measured Raman spectrum $\boldsymbol{D}$ of a solution is modelled as the linear combination of the solvent spectrum $\boldsymbol{S_B}$, the SC spectrum $\boldsymbol{S_{SC}}$, and a residual error matrix $\boldsymbol{E}$. The assumption of linearity applies for dilute solutions. In the two-component system, the Raman-MCR decomposition can be written as:

$$\boldsymbol{D} = C_B\boldsymbol{S_B} + C_{SC}\boldsymbol{S_{SC}} + \boldsymbol{E} \quad \textbf{(1)}$$

where $C_{SC}$ and $C_B$ represent the fractions of the components. To compute $\boldsymbol{S_{SC}}$, we take the non-negative minimum area difference between $\boldsymbol{D}$ and $\boldsymbol{S_B}$,which can be measured with high precision. Here we perform Raman-MCR on mesoscale oil droplets in water for the first time, which has several advantages. First, because the emulsion is a solution, we measure an ensemble of droplets. Second, the high surface to volume ratio of oil droplets provides far more interfacial area than planar interfaces. Third, the SC spectra of the interface can be directly compared with the Raman spectra of the bulk solvent, unlike in VSFG or VSFS.

Figure 1(a,b) shows our in-house confocal Raman microscope used for SC measurements on both small molecules and oil droplets. We validated our setup on small molecule alcohol solutions of methanol, ethanol, propanol, and butanol (Supplementary Fig. 1), obtaining SC spectra with the same shape and magnitude as in prior reports (Supplementary Fig. 2). This is also an important reference to show how a molecular interface differs from the mesoscopic scale of the oil-water emulsions, as will be seen below. We then prepared the oil droplet emulsions from regular and fully deuterated n-hexadecane ($C_{16}H(D)_{34}$) in neat water ($H_2O$) and heavy water ($D_2O$) with ultrasonication in cleaned glassware (see Methods, Supplementary Note 2, and Supplementary Fig.

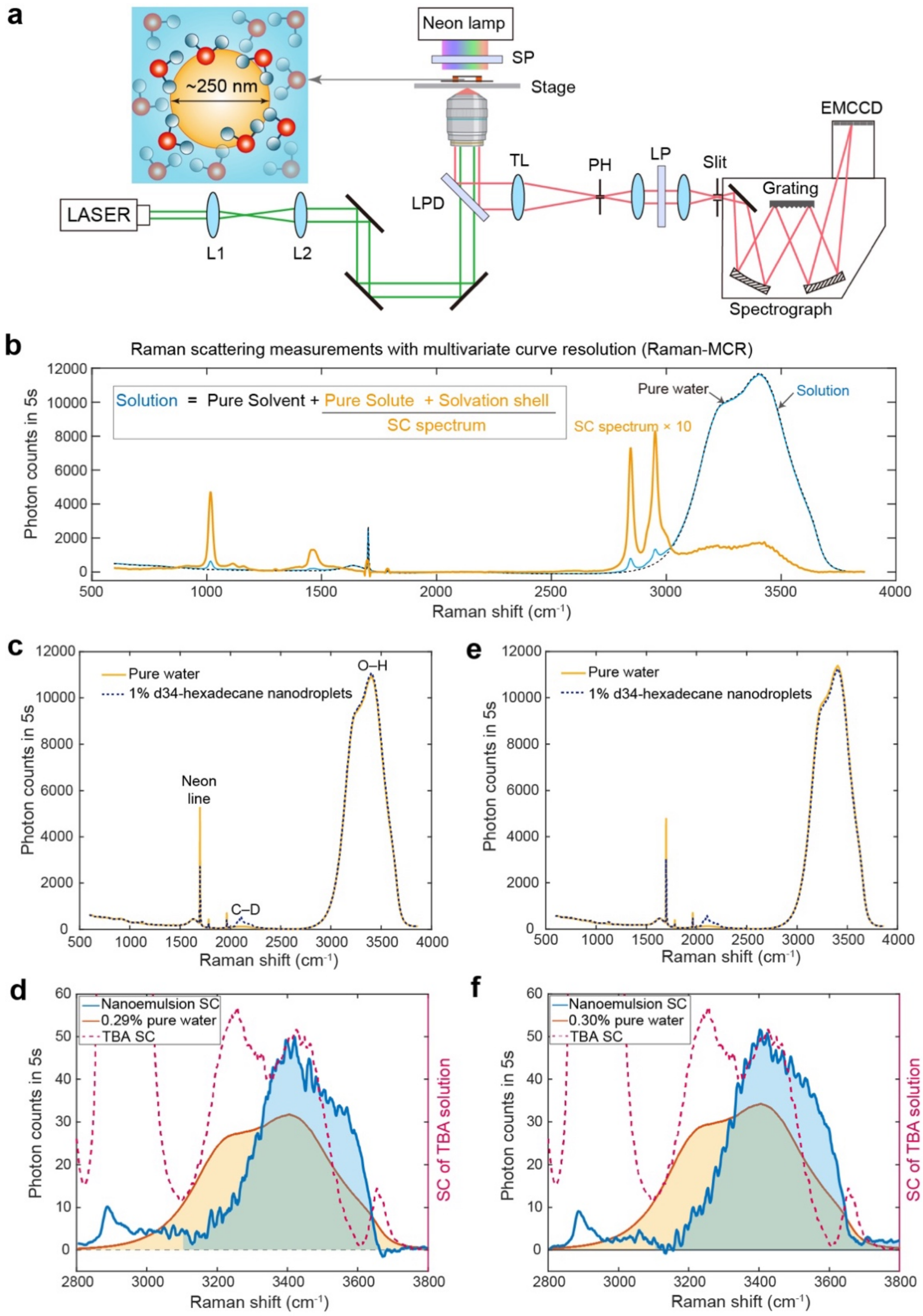

**Figure 1. Raman-MCR Spectroscopy on oil droplets.** (a) Schematics of home-built confocal micro-Raman system. L: lens; LPD1, LPD2: long-pass dichroic beam splitter; SP: short-pass filter; LP: long-pass filter. The set-up is featured by under-filling the back aperture of a 20× objective lens (effective numerical aperture N.A. ~ 0.4) and still capturing the scattered Raman signal with a large collection efficiency. This configuration yields a confocal volume of ~ 2 µm × 2 µm × 10 µm. An optical pinhole size of 40 µm is chosen based on the first minimum of the Airy disk from a point source. This size is suitable to block the scattering effect of oil emulsions while keep sufficient photon counts (Supplementary Fig. 5). (b) Schematic illustration of Raman hydration-shell spectroscopy. (c, e) Two independent measurements of 1% d34-hexadecane oil emulsion in water using Raman-MCR spectroscopy. (d, f), SC spectra of oil emulsions (blue solid) and 0.5 molar tertiary butyl alcohol (TBA) (dashed) in water, area-scaled Raman spectra of pure water (orange solid).

3). The diameter of the oil droplets is ~0. 25 $\mu$m with a narrow distribution (polydispersity index (PDI)<0.3; Supplementary Fig. 4). The emulsions are stable over weeks to months.

However, performing Raman-MCR on oil droplets is more challenging than performing it on small molecules such as alcohols because, despite providing more interfacial surface area than planar interfaces, they provide less interfacial area than small solutes. Thus, acquiring a true SC spectrum requires an exceptionally high signal-to-noise ratio (usually >1000:1 to 10000:1[31]) and strict spectral stability (less than 0.5 cm$^{-1}$). A high SNR typically requires a long acquisition time, over which maintaining spectral stability is necessary but difficult. Here we address this challenge by installing a Neon lamp on the microscope. The narrow emission line from Neon lamp serves as a "beacon" with absolute energy (Supplementary Fig. 6). This enabled the final noise floor of the control to be pushed to a negligible level as shown in Supplementary Fig. 7.

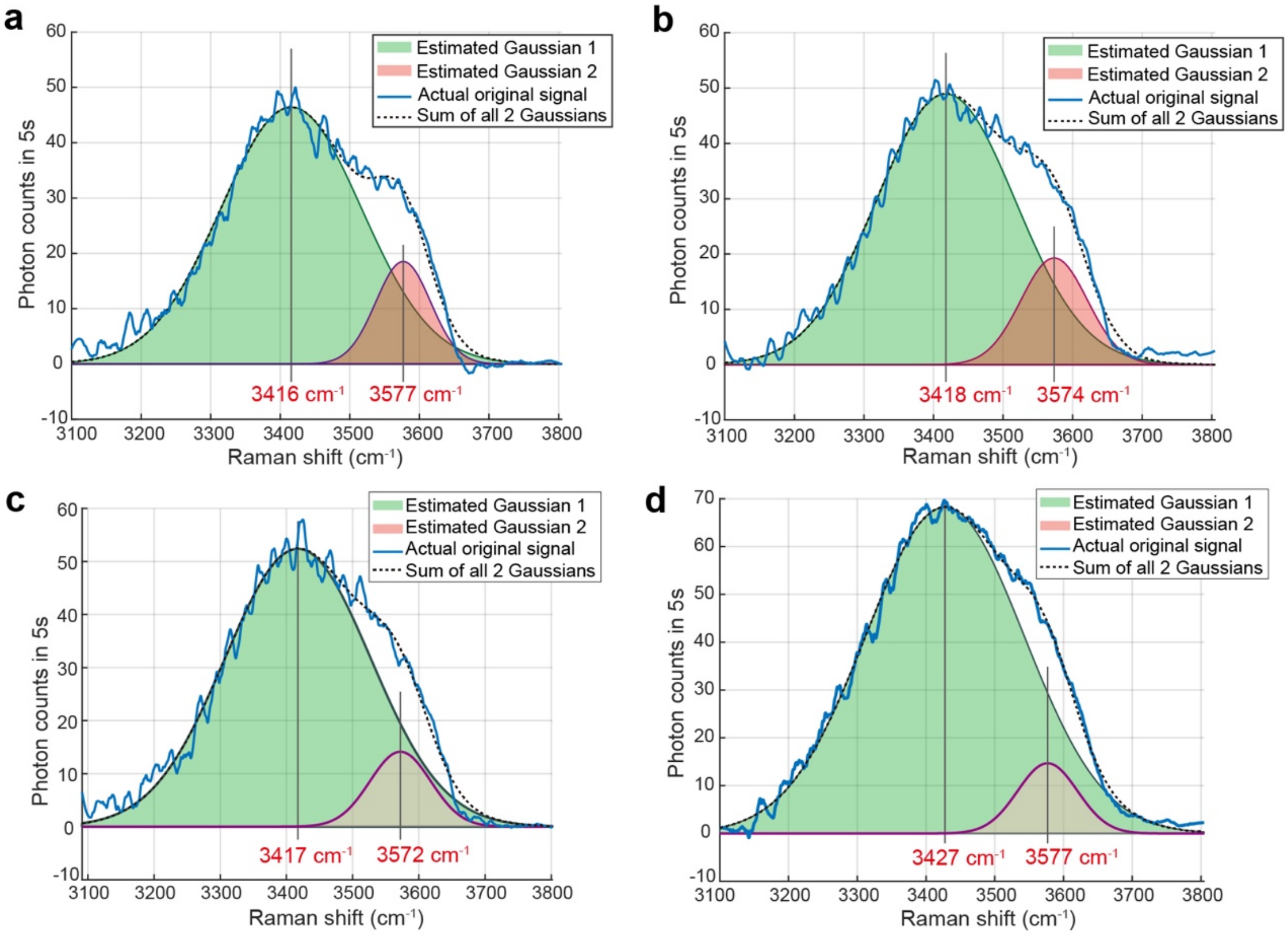

**Figure 2. Oil-perturbed, solute-correlated (SC) water spectra of oil droplets**. SC spectra (blue) fitted by Gaussian profiles for (a,b) replicates of 1% d34-hexadecane droplets in water, (c) 2% regular hexadecane droplets in water, and (d) 2% d34-hexadecane droplets in water. The percentage of the estimated gaussian 2 component over sum is 13.5%, 15.8%, 10.2% and 7.5% for (a-d), respectively.

We then measured Raman-MCR spectra for fully deuterated 1% n-hexadecane ($C_{16}D_{34}$) droplets as seen in Figure 1(c,e). We use deuterated oil because the solute's CD-stretch band is spectroscopically separated from the OH-stretch band of water. After applying the spectral correction, we obtained robust measurements of the SC spectrum of the oil droplets as provided in Figure 1 (d,f); the area under curve of the SC spectrum is 0.29% and 0.30% of that of the bulk water's OH-stretch peak for two replicates. We then fit the OH-stretch region of the SC spectra with a minimum number of Gaussian profiles, and we obtained consistent results: two sub-bands peaked at 3416 cm$^{-1}$ and 3577 cm$^{-1}$ in one replicate (Figure 2a), and at 3418 cm$^{-1}$ and 3574 cm$^{-1}$ in the other replicate (Figure 2b).

We also prepared and measured other oil emulsions, including 2% d34-hexadecane droplets in water and 2% hexadecane droplets as shown in Figure 2(c,d). The corrected SC spectra yields an area ratio of 0.40% that of bulk water for the former and 0.28% that of bulk water for the latter. We also fit the OH-stretch region of the SC spectra for these samples, determining sub-bands peaked at 3417 $cm^{-1}$ and 3572 $cm^{-1}$ for d34-hexadecane (Figure 2c), and at 3427 $cm^{-1}$ and 3577 $cm^{-1}$ for hexadecane (Figure 2d). The consistent measurements between both technical replicates and different samples verify the robustness of our method.

**Interfacial water structure at oil droplets**

Unlike with VSFG and VSFS, we can directly compare the SC spectra to the spectra of bulk water or that of solutions of small solutes such as TBA (Figure 1(d,f) and Figure 2), enabling us to gain unprecedented insight. This direct comparison shows substantial differences between the spectra, with two notable differences in particular. First, besides the major 3420 $cm^{-1}$ peak in the SC spectrum, which is close to the main peak of the bulk water, a new high frequency sub-band peaked at ~3575 $cm^{-1}$ is observed in every droplet sample. This peak is remarkable for two reasons: its area is substantial, accounting for about 16-18% of the entire SC spectrum, and its frequency of ~3575 $cm^{-1}$ has not been previously documented in similar systems.

Secondly, the shoulder peak at 3250 $cm^{-1}$ has nearly vanished in all SC spectra of the oil droplets (Figure 1(d,f) and Figure 2). This is in contrast to the SC spectra of small oily solutes[29,33-35] for which this shoulder peak increases in prominence compared to bulk water (Figure 1(d,f); Supplementary Fig. 2e-h). In general, this peak has been linked to strong hydrogen bonding between adjacent water molecules, which weakens the covalent OH bond and causes the O-H vibration to shift towards lower frequencies (i.e., redshift). In particular, it has been linked to tetrahedral order in the local hydrogen bonding network[41-43] because tetrahedral order can facilitate strong hydrogen bonding. Recent work attributes the shoulder's fundamental origin to Fermi resonance between the OH-stretch modes and the bending overtone[36,44,45] and attributes the link between the shoulder and tetrahedral order to the fact that strong hydrogen bonds are required to the redshift the OH-stretch frequencies down to the bending overtone. Hence, the decline of this peak in the measured SC spectra suggests that the hydrogen bonding network is weakened and more disordered at the mesoscopic oil/water interface, in contrast to that found with small molecule hydrophobic interfaces[29,33-35]. Correspondingly the Raman-MCR spectrum of the oil phase shows no shift in frequency arising from interfacial effects (Supplementary Fig. 8), unlike that found for VSFS in which a 5-10 $cm^{-1}$ shift was reported[11], but instead shows small structural perturbations from the water arising from the known increases in oil ordering at the interface[46-48].

To gain theoretical insight into the interfacial water structure, we next conducted MD simulations of an oil-water system in a periodic box. We modelled the oil layer as dodecane (similar to hexadecane used experimentally) employing the OPLS model[49] and the water layer using the SPC/E model[50], which, despite being a fixed-charge water model, captures some important features of water-organic interfaces, such as the distribution of free OH groups[51]. We use the Willard-Chandler interface (WCI) to characterize how distant a water is to the interface[52]. We show the liquid density as a function of distance to the WCI in Figure 3(a), and then calculate the degree of order as a function of distance using two variants of the Errington-Debenedetti order parameters[53], $q_4$, which considers the four nearest neighbors around a central oxygen, and a version using only the nearest three neighbors as $q_3$, to measure the degree of tetrahedral ordering in the fluid and at the interface (Supplementary Information). We show their average values (denoted as $\overline{q_4}$ and $\overline{q_3}$) as a function of distance to the WCI in Figure 3(b,c). While $\overline{q_3}$ and $\overline{q_4}$ resemble bulk water at 298 K at distances greater than around 4 Å into the liquid layer, they decline substantially at distances closer to the interface, indicating that the tetrahedral nature of the hydrogen bonding network is reduced at the oil-water interface. This confirms our qualitative interpretation of the reduced 3250 cm$^{-1}$ peak of Raman-MCR experiment.

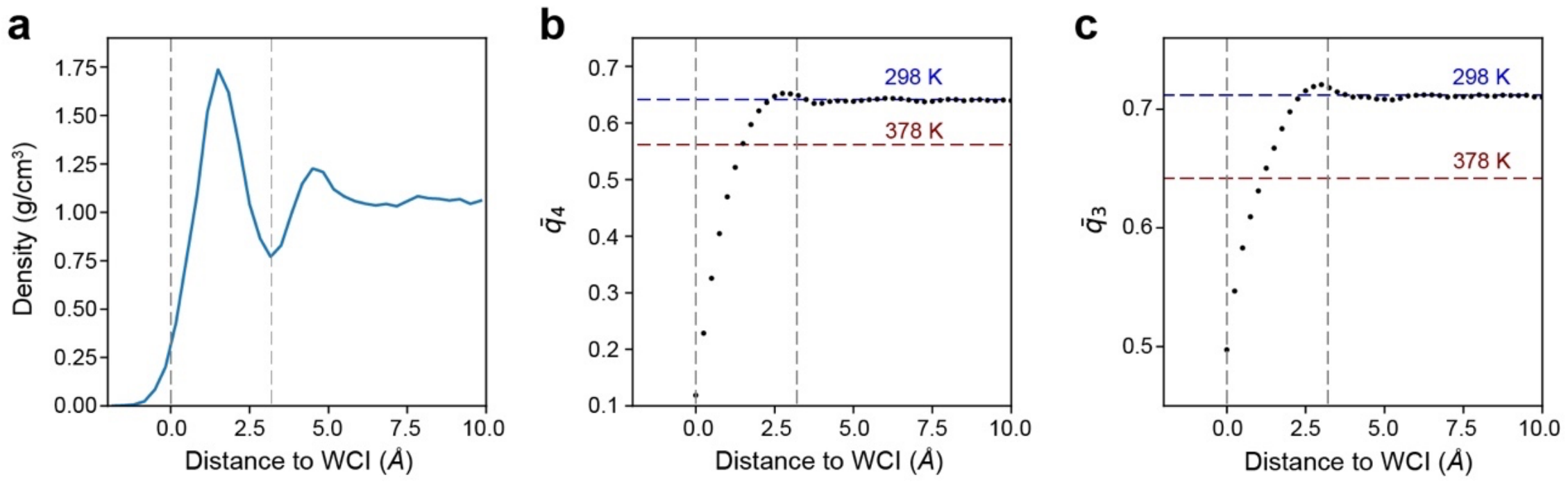

**Figure 3. Tetrahedral order of water at water/oil interface.** (**a**) Liquid density as a function of distance to the WCI for our MD simulation of a water/dodecane system. Positive distances correspond to the water side of the surface. We show the average values of order parameters (**b**) $q_4$ and (**c**) $q_3$ for waters as a function of distance to the WCI together with the average value for bulk water at two different temperatures. The region of the first peak in the WCI is demarcated with two vertical grey lines. See the supporting information for complete details on the MD simulations and calculation of the order parameters.

We then probe whether the hydrogen bonding network is weakened and thus exhibits a decline in the 3250 cm$^{-1}$ shoulder using our recently developed monomer-field model for Raman spectroscopy[36]. This model considers how the potential energy surface of individual water monomers is deformed by the electric fields produced by neighboring water molecules[36]. The electric fields experienced by the hydrogens, projected along the OH-stretch, are the only environmental input to the monomer-field model, and they reflect the strength of the hydrogen bonding between waters. As seen in Figure 4a, the electric fields experienced by bulk water are close to a unimodal distribution with moderate-to-strong field strength, indicating most of the OH groups are hydrogen bonded. We also show the fields experienced by waters within 2 Å of the interface, divided into two distinct populations: waters without (~75%) and with (~25%) a free OH. The electric fields experienced by waters without free OH's (i.e., both OH groups are participating in hydrogen bonds) resemble those experienced by bulk water except that they are somewhat weaker. This decrease directly indicates that hydrogen bonding is weaker for these waters, which parallels the reduced tetrahedral order observed in Figure 3(a,b). In contrast, the distribution of electric fields experienced by waters with a free OH is bimodal, consisting of one peak at much lower field strengths, which corresponds to the free OH, and another peak at slightly higher field strengths than the bulk liquid, which corresponds to the hydrogen-bonded partner OH-stretch for waters with a free OH. When considering both populations of water together, they have weaker average electric fields than in the bulk liquid.

We next use our monomer-field model to produce a computational SC spectrum to compare directly with experiment. For the resulting interfacial spectrum in Figure 4b, we used the electric fields for all waters within 10 Å of the WCI. The shoulder at 3250 cm$^{-1}$ in the computational spectrum is substantially reduced relative to the bulk spectra, paralleling the same reduction seen in the Raman-MCR experiment, although we do not observe the complete loss of Fermi resonance as in experiment. Such consistency between the experiment and the theory in regards the reduced 3250 cm$^{-1}$ peak supports our conclusion of lowered tetrahedral order and a weakened hydrogen bonding network.

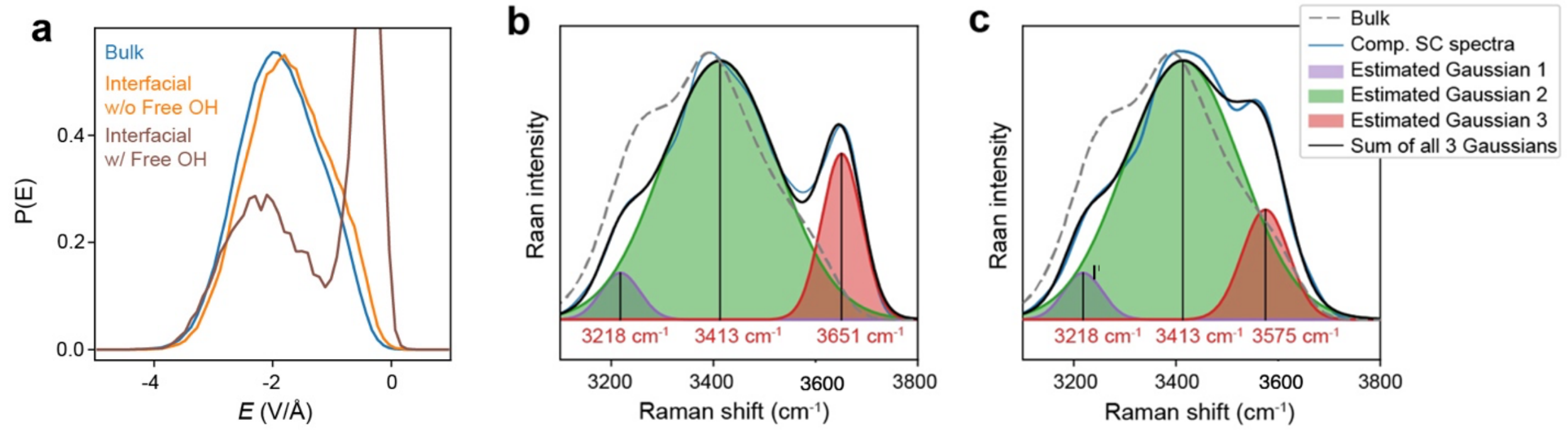

**Figure 4. Connecting intermolecular electric field to Raman spectra using the monomer-field method**. (a) The electric field (*E*) distribution measured within 2 Å of the WCI, dividing the measured distribution into contributions from waters with and without a free OH. We only examine waters within 2 Å of the WCI because the Raman SC spectrum becomes bulk-like when considering waters>2 Å (Supplementary Fig. 9). (b) Computational SC spectra, whose free OH peak is at 3651 $cm^{-1}$. The shoulder at 3250 $cm^{-1}$ is significantly reduced relative to the main peak near 3400 $cm^{-1}$. (c) Computational SC spectra with an additional electric field of -67.5 MV/cm applied along the free OH due to the zeta potential.

## Electric fields at the interface of oil droplets in water

We next address the emergence of the new sub-band at ~3575 $cm^{-1}$ arising from the free OH stretching mode. While it is observed in every droplet sample (Figure 2), it is not reproduced by the simulated SC spectrum (Figure 4b) from the monomer-field model which predicts a peak around 3650 $cm^{-1}$. Yet several considerations all suggest that this 3650 $cm^{-1}$ peak should be fundamentally linked to the new 3575 $cm^{-1}$ one. First, both the amplitude and width of this 3650 $cm^{-1}$ peak are comparable to that of the 3575 $cm^{-1}$ found experimentally, implying a common origin. Second, we find in our simulations that a substantial ~25% of waters within 2 Å of the WCI have a dangling OH group, indicating that ~12.5% among of all interfacial OH groups are dangling, which matches the ~8-16% relative area of 3575 $cm^{-1}$ peak in the experimental SC spectrum (Figure 2). Finally, previous work on water in contact with oils typically finds a free OH-stretch around 3650 – 3670 $cm^{-1}$, including the SC-spectra of TBA and for other alcohols in Supplementary Fig. 2 and Supplementary Note 3, as well as gas phase measurements[54]. Thus given the increased population of water with a free OH at the interface, a substantial peak at 3650 $cm^{-1}$ would be expected.

But what makes oil-in-water emulsions different from a planar interface or the small molecule solutions is the zeta potential due to a net negative charge on the oil droplet, whose presence is irrefutable but whose fundamental origin remains unknown[15,24,44]. While we cannot resolve the controversy over the origin of the zeta potential, we can assume that it would create an extra electric field emanating from the oil droplet – i.e. beyond what the OH-stretch experiences simply by being in the vicinity of an alkane or other water molecules – that further redshifts the free O-H frequency by 80 $cm^{-1}$ from 3650 $cm^{-1}$ to ~3575 $cm^{-1}$. Using the vibrational Stark experiment that assumes the OH peak is linearly dependent on the local electric field (EF), the slope, which is called the Stark tuning rate $|\Delta\vec{\mu}|$, has been calculated from small water clusters as $2.4\ \text{cm}^{-1}/(\text{MV/cm})$[55]. This number is very close to $|\Delta\overrightarrow{\mu_{OH}}|$ of the OH stretch modes of phenol, which has been measured by vibrational Stark spectroscopy to be $2.6 \pm 0.3\ \text{cm}^{-1}/(\text{MV/cm})$[56]. Therefore, an 80 $cm^{-1}$ shift observed experimentally represents an EF = 80 $cm^{-1}$/2.4 $cm^{-1}$/(MV/cm) ~30 MV/cm. Alternatively, Geissler et. al. found a tuning rate of $1.6\ \text{cm}^{-1}/(\text{MV/cm})$[57,58], which yields an EF = $80\ \text{cm}^{-1}/\ 1.6/(\text{MV/cm}) \sim 50\ \text{MV/cm}$. We previously reported fields of ~8.5

MV/cm at the interface of water microdroplets using a nitrile bond of an organic fluorophore as a local probe[59,60]. However, the probe's size is likely comparable to or even larger than the scale of the interface (1-2 layers of water), so the reported electric field may be underestimated in that study.

Simulation can also provide crucial insights. The SC spectra simulated in Figure 4b only accounts for a neutral droplet with no net negative zeta potential. Hence we adapt the Raman monomer-field model to now include this additional electric field introduced by the oil. We validate our monomer-field calculation with the extra electric field by asking 1) how large of an EF is required to observe the redshift, and 2) whether the computational SC spectrum resembles the experimental SC spectrum. We adopt the simple assumption that the electric field from oil acts on all free OH stretches with equal intensity. This is justified due to the bimodality of water EFs in Figure 4a in which the free OH stretch is much weaker and thus more sensitive to the extra oil electric field than its hydrogen-bonded OH partner on the same molecule. As show in Figure 4c, we find that a field strength of -67.5 MV/cm results in an OH peak at ~3575 $cm^{-1}$, which reasonably close to the values of 30 - 50 MV/cm found from the Stark shift estimation of the experimental measurement. Despite the simplicity in our modelling of the extra oil electric field arising from the zeta potential, our final simulated SC-spectrum nearly quantitatively resembles the SC spectra obtained experimentally. Not only are the frequencies of the sub-bands reproduced, but also their relative amplitudes, thus lending support to the large influence of the oil droplet charge on observed electric fields at the hydrophobic-water interface.

## Discussion and Conclusion

While water at hydrophobic interfaces has long been recognized to play a fundamental role in nature, measuring interfacial spectra is difficult and controversial (Supplementary Note 1). Raman-MCR spectroscopy offers "in-solution" interface selectivity, which has proven useful for investigating the solvation shell of small molecules in aqueous solution. Herein, we combine the reliability of carefully prepared oil-water droplet interfaces with the robustness of Raman-MCR spectroscopy, bypassing some of the challenging limitations of other surface-specific spectroscopic methods. Our experimental results are highly reproducible between replicates and consistent across multiple oil-in-water samples. Our theoretical work on Raman spectroscopy is also timely in that it allows us to extend it to understanding the underlying features of the SC spectra for these interfacial systems. The combination of measurements and theory provide mutual support and complementary insight into the molecular nature of oil droplets in water.

The use of Raman-MCR approach offers an intrinsic advantage in that its result can be readily compared to other solution phase measurements including bulk water and solutions of small hydrophobic solutes. Note that this important comparison is not shared by previous vibrational sum frequency techniques which cannot generate signal in bulk water or solution of molecular solutes (much smaller than the wavelength of light). One of the key insights is that while the molecular hydrophobic interface exhibits an enhanced tetrahedral order compared to bulk water[29,33-35], which was also reproduced in our own measurement, the mesoscopic interface of the oil droplets (~ 0.25 $\mu$m size) displays clearly reduced tetrahedrality and a weakened hydrogen bonding network. This qualitative transition is in line with the crossover from the small to large lengthscale of water hydration physics near small verse large hydrophobic solutes[29,61,62].

In contrast to the weakened hydrogen bonding network, the local electrostatics is found to be greatly enhanced at the oil interface. The Raman-MCR shows a consistent redshift of the free OH peak by ~80 $cm^{-1}$ relative to that found at gas phase and small solutes in water (Supplementary Note 3). We estimate from the experiment and theory that the free OH peaks experience electric fields with magnitudes of ~35-70 MV/cm. We suspect that this electric field is related to the negative zeta potential associated with oil droplets[15,23,24], whose molecular origin remains unclear. A prior VSFS study suggested the existence of C−H···O−H improper hydrogen bonds at the water-oil droplet interface which encourages charge transfer that might explain this red-shift.[11] However it is difficult to see how this leads to such a large shift in the free OH peak, since the OH stretch frequency is much more sensitive to the hydrogen's local environment than that of the oxygen (*e.g.*, in the water dimer, the donor OH stretch is more redshifted than in the receptor OH stretches[63]). Another possibility is the increased presence of hydroxide ions at the water-hydrophobic interface[10] that can give rise to organized electric fields of ~50 MV/cm[64]. Finally, contact electrification is also a possible scenario. Although its molecular nature is still not completely explained[65], contact electrification has recently been demonstrated to occur in various water-related systems including liquid-liquid (such as oil/water) interfaces[66-68].

Nonetheless, regardless of the exact origin of the zeta potential, we were able to successfully reproduce the experimental SC spectrum by assuming that an electric field of ~35-67 MV/cm is acting only on the free OH stretches. This implies that the electric field associated with the zeta potential is strongest at the free OH and does not extend much further below the interface, and that the hydrogen-bonding partner of the free OH water is more immune to the oil EF perturbation. In this sense, the structural weakening of the hydrogen bonding network is necessary for the OH stretch to respond to the extra oil electric field. Hence, our key finding is that, water structural weakening and enhanced electrostatics together are emergent features at the mesoscale interface of oil-water emulsions, distinguishing it from planar interface or molecular hydrophobic interfaces.

Such a strong electric field would greatly impact chemical reactivity. Assuming a difference dipole moment between the transition state and the reactant ground state of 2 Debye, which is a reasonable magnitude for many reactions[69], the ~50 MV/cm average electric field would lower the free energy barrier by 4.8 kcal/mol. Using simple transition state theory[70], this would accelerate the reaction rate by over 3,000 times at room temperature. This is likely a key factor for explaining the emerging field of water microdroplet chemistry in which the rate acceleration of various reactions has often be measured to be in the range of $10^3$ to $10^6$ compared to bulk solvent[37,39,40,71,72].

## Acknowledgement

We acknowledge the support of the Multi-University Research Initiative (MURI) of the Air Force Office of Scientific Research (FA9550-21-1-0170). W.M. acknowledges support from the National Institute of Health (R01 EB029523). T.H.G, R.A.L. and J.P.H. also acknowledge the support of DOE CPIMS for the monomer-field theory applied to this work.

## Author contributions

L.S. X.L. and W.M. developed the new experimental approach, R.A.L, J.P.H. and T.H.-G. formulated all the theoretical work. L.S., R.A.L, T.H.G. and W.M. wrote the manuscript, and all authors contributed to all insights through extensive discussion.

## Competing interests

All authors declare no competing interests.

## Data Availability

Source data for Figures 1-4 are available with this manuscript.

## Code Availability

All analysis scripts are in a private github repo but are available upon request.

# Supplementary Information

# Water Structure and Electric Fields at the Interface of Oil Droplets

Lixue Shi[1,2†], R. Allen LaCour[3,4†], Xiaoqi Lang[1], Joseph P. Heindel[3,4], Teresa Head-Gordon[3,4*], Wei Min[1*]

[1]Department of Chemistry, Columbia University, New York, NY, 10027

[2]Shanghai Xuhui Central Hospital, Zhongshan-Xuhui Hospital, and Shanghai Key Laboratory of Medical Epigenetics, International Co-laboratory of Medical Epigenetics and Metabolism, Institutes of Biomedical Sciences, Shanghai Medical College, Fudan University, Shanghai 200032

[3]Kenneth S. Pitzer Theory Center and Department of Chemistry, University of California, Berkeley, CA, 94720 USA

[4]Chemical Sciences Division, Lawrence Berkeley National Laboratory, Berkeley, CA, 94720 USA

[†]Authors contributed equally

*E-mail: **wm2256@columbia.edu** and **thg@berkeley.edu**

## MATERIALS AND METHODS

### Chemicals:

Hexadecane (52209, Sigma, analytical standard), d34-hexadecane (489603, Sigma, 98 atom % D), sulfuric acid (Sigma), $H_2O_2$ (Sigma), $D_2O$ (Cambridge Isotope) were used as received. All glassware (glass vials for oil droplets preparation, glass coverslips, glass slides) was first cleaned by soaking in a freshly prepared piranha (3:1 $H_2SO_4$:$H_2O_2$) solution for 45 min. Then, glassware was further rinsed by large amount of ultrapure Milli-Q water for 5-10 times to thoroughly remove piranha residual. Glass vials and slides were dried in 90°C oven and cool to room temperature before use. Glass coverslips were air dried before use.

### Preparation of oil droplets in water

1-2 vol% hexadecane was added to pure $H_2O$ or $D_2O$ in a cleaned glass vial. The solution was first mixed for ~3 min using a vortex and then placed in an ultrasonic bath until monodisperse droplets were formed (no longer than 15 minutes). After size and zeta-potential characterization, oil droplets solutions were sandwiched between a cleaned coverslip and a cleaned microscope slide using an 8-mm silicone spacer (JTR8R-A2-0.5, 8-9 mm DIAx0.8 mm Depth, PSA on both sides) for Raman micro-spectroscopy.

**Measurement of the size and zeta-potential of oil droplets in water**

The size distribution of the oil droplets was measured with dynamic light scattering (DLS) using a Malvern ZS Nanosizer instrument. Specifically, the oil emulsions were diluted in either $H_2O$ or $D_2O$ to 0.02 vol. % for measurement. Zeta-potential $\zeta$ values were calculated from the electrophoretic mobility values using the following expression: $\mu = \frac{\varepsilon\varepsilon_0}{\eta}\zeta f(\kappa a)$, where $\varepsilon_0$ is the vacuum permittivity, $\varepsilon_0$ is the relative permittivity of water, $\eta$ is the viscosity of water, $f(\kappa a)$ is Henry's function. In the software of the DLS instrument, Hückel formula was used as $f(\kappa a) \rightarrow 2/3$. All measurements were performed at 25 °C with disposable plastic cuvettes. All the samples studied in this work had similar size distribution with z-average diameters in the range of 200-400 nm, with a polydispersity index (PDI) < 0.3. Calculated zeta-potential values are in the range -55 mV < $\zeta$ < -65 mV.

**Raman spectroscopy**

All Raman spectra were acquired with a home-built confocal Raman micro-spectroscopy. A 532-nm laser (Samba 532 nm, 400 mW, Cobolt Inc.) is used as the light source. The laser beam was first collimated and expanded by telescope lenses [f1=40 mm, f2=300 mm, Thorlabs] to achieve the designed illumination spot size. The expanded beam was then directed to an inverted microscope (IX71, Olympus) installed with a dichroic beam splitter (LPD1, LPD02-532RU-25, Semrock). A microscope objective (UplanSApo 20×, N.A. = 0.75, Olympus) was underfilled and the resulting illumination spot had a dimension of ~ 2 μm × 2 μm × 10 μm. The emitted Raman signal first passed a pinhole (40 μm, Thorlabs) for background suppression and was relayed by two lenses (Thorlabs) before being projected to the spectrometer (Kymera 328i, Andor) that is equipped with a grating of 600 lines/mm, blazed at 500 nm, allows a bandpass of 121 nm and a spectral resolution of 7.6 $cm^{-1}$. A long-pass filter (LP03-532RU-25, Semrock) was installed between the relay lenses to block laser light from Rayleigh scattering. Raman signal was then collected by an EMCCD with an active pixel-area of 200×1600 (Newton970, Andor). All the data collection was performed with a custom LABVIEW program (National Instruments). Prior to the preprocessing procedures, the spectra were calibrated by 1:1 mixture of acetonitrile and toluene taken on the same day.

We implemented spectral corrections with narrow emission peaks with absolute energies (Supplementary Fig. 6). To perform sub-pixel spectral correction, a Neon lamp (Thorlabs CSL1) was set up on the stage with a 600-nm SP filter (Thorlabs FES0600). The emitted line around 1700 $cm^{-1}$ (585.25 nm) was fitted by a Gaussian shape for each of the scan. The frequency shifts among different scans of a sample and between different samples were corrected by precisely match the fitted Neon peak position to 1699.5 $cm^{-1}$. The interested region of the raw spectrum was first fitted with a spline curve using a sub-pixel step size (d) of 0.1 cm-1. The shifts were performed as wave1=wave0(p+d*N), where N is the number of sup-pixel steps, and wave1 and wave0 are the shifted and un-shifted solution spectra, respectively. The spectral fitting and shifting were performed with a MATLAB program. To reduce noise on oil droplet solutions, Savitzky-Golay smoothing (second-order, 275-point) was performed on the SC spectra. Alcohol solutions were measured in glass-bottom dishes with 200-mW on panel (50-mW on sample), 5s integration time per run, averaged over 100 runs.

**Spectral Corrections**

In prior work, spectral corrections with Neon or Xeon lamps have been utilized to compensate sub-pixel wavelength shifts due to barometric pressure changes that could lead to spurious features upon spectral decomposition[1].We first calibrated the stability of our system with Neon emissions by performing continuous spectral scanning with a 5s integration time for each scan. Considering a fitting error of the Neon emission peak of ~0.1 $cm^{-1}$, we expect the residual SC background after correction can be suppressed to below 0.1% of pure water. Also, we performed a self-checking procedure to confirm the instrumental stability: we perform a “water-subtract-water” negative control first on the pure water sample to check if corrected residual background is smaller than 0.1% of pure water (Supplementary Fig. 7).

**MD Simulations**

We used the Lammps simulation package[2] to model the water-oil interface. We modelled the interface using a slab geometry, as is common. We modelled the water using the SPC/E model[3] and the oil layer using the long-chain variant of the OPLS model[4]. We used 360 oil molecules and 6804 water molecules. We truncated interactions at 15 Angstroms. Long-range electrostatics were evaluated using a particle-particle particle-mesh solver[2] with a grid size of $10^{-6}$. Simulations were kept at a constant temperature of 298 K and a constant $P_{zz}$ of 1 atm, where $P_{zz}$ is the component of the pressure tensor perpendicular to the interface. The simulation was run for approximately 5 nanoseconds with a time step of 0.5 fs.

We examined the $q_4$ order parameter:

$$q_4 = 1 - \frac{3}{8}\sum_{j=1}^{3}\sum_{k=j+1}^{4}\left(\cos\Psi_{ijk} + \frac{1}{3}\right)^2$$

and, because many waters at the interface have fewer than four neighbors, we examined a three neighbor variant:

$$q_3 = 1 - \frac{6}{8}\sum_{j=1}^{2}\sum_{k=j+1}^{3}\left(\cos\Psi_{ijk} + \frac{1}{3}\right)^2$$

For both order parameters, $\Psi_{ijk}$ is the angle between a central water oxygen $j$ and each pair $i$ and $k$ of its $n$ nearest water neighbors. Both quantities take on a value of 1 in a perfect tetrahedral environment and a value of 0 in an ideal gas. Because we wish to compare order between the bulk, where most waters have at least three neighbors, and the interface, where many waters have less than three neighbors, we follow Hande et al. and only evaluate $q_n$ for waters with $n$ neighboring oxygens within a specific distance of the central water.[5] However, we use a cutoff of 3.5 Å, which is much smaller than the 6.3 Å cutoff used by Hande et al.

**Monomer Field Model for Raman Spectroscopy**

With the monomer-field model, we first measure the pairs of electric fields ($E_1$ and $E_2$) experienced by each hydrogen on a water molecule projected along the corresponding OH-stretch. For a random subset of the measured $E_1$ and $E_2$), we then use a variational calculation the Schrodinger equation on a field-dependent energy surface representing a water monomer in the electric fields produced by neighboring water molecules. The electric field-dependent energy surface is given by:

$$U_{liq}(E_1, E_2) = U_{mono} + U_{stretch}(E_1, E_2) + U_{bend}(E_1, E_2)$$

where the $U_{mono}$ is the gas-phase monomer energy surface and $U_{stretch}$ and $U_{bend}$ represent the perturbations due to neighboring molecules. The $U_{stretch}$ and $U_{bend}$ terms were parameterized to reproduce previously known relations for the stretching frequencies of water. Solving the Schrodinger equation yields a set of frequencies for each pair of $E_1$ and $E_2$. We replace each frequency with a Gaussian distribution having a full-width at half-max of 70 $cm^{-1}$ and a height proportional to the frequency's Raman activity, which we evaluate using a 1-body polarizability surface. The sum of the Gaussian distribution is our predicted Raman spectrum. We use the same energy and polarizability surfaces used in our previous work[6]. In Figure 4c we add an extra field to each water with a free OH-stretch. We evaluated whether a water had a free OH stretch by examining whether the OHO angle formed between a water and any neighbor within 3.5 Å was less than 90°.

## SUPPLEMENTARY NOTES

### Supplementary Note 1. A literature overview on the structure of interfacial water by second-order optical spectroscopy

Contradictory conclusions have been drawn about the structure of water at hydrophobic interfaces, as different shapes of spectra have been reported over three decades of literature on nearly the same sample. A similar assessment was made in *Science* **2021,** *374* (6573), 1366-1370.

| Year | Group | Sample | Technique | Conclusion | Spectra | Ref. |
|---|---|---|---|---|---|---|
| 1994 | Shen | water/hexane | Reflection SFG | hexane/water and air/water interfaces have similar water ordering | 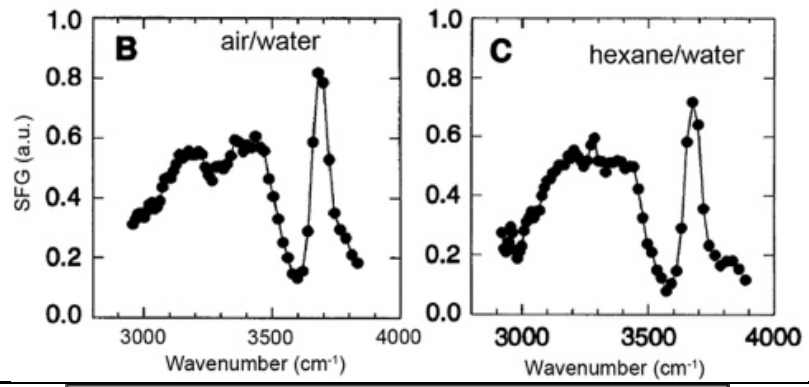 | *Science* **1994,** *264* (5160), 826-828 |
| 2001 | Richmond | hexane/ water; CCl4/water | Reflection SFG | water at CCl4/water and hexane/water interfaces are weaker compared to the air/water interface | 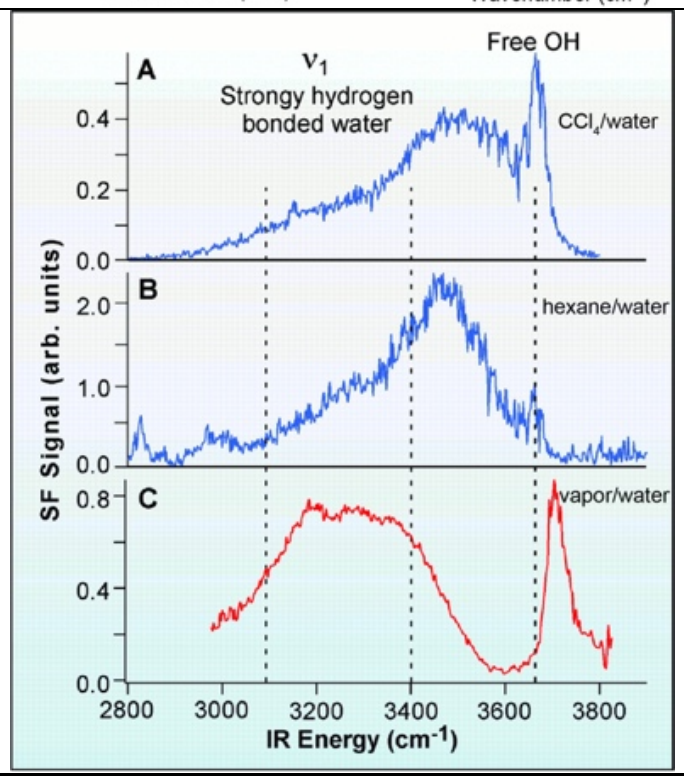 | *Science* **2001,** *292* (5518), 908-912 |
| 2003 | Richmond | alkane/ water | Reflection SFG | alkane/water interface has weaker H bond compared to air/water; but a more intense non-hydrogen bonded O-H peak | 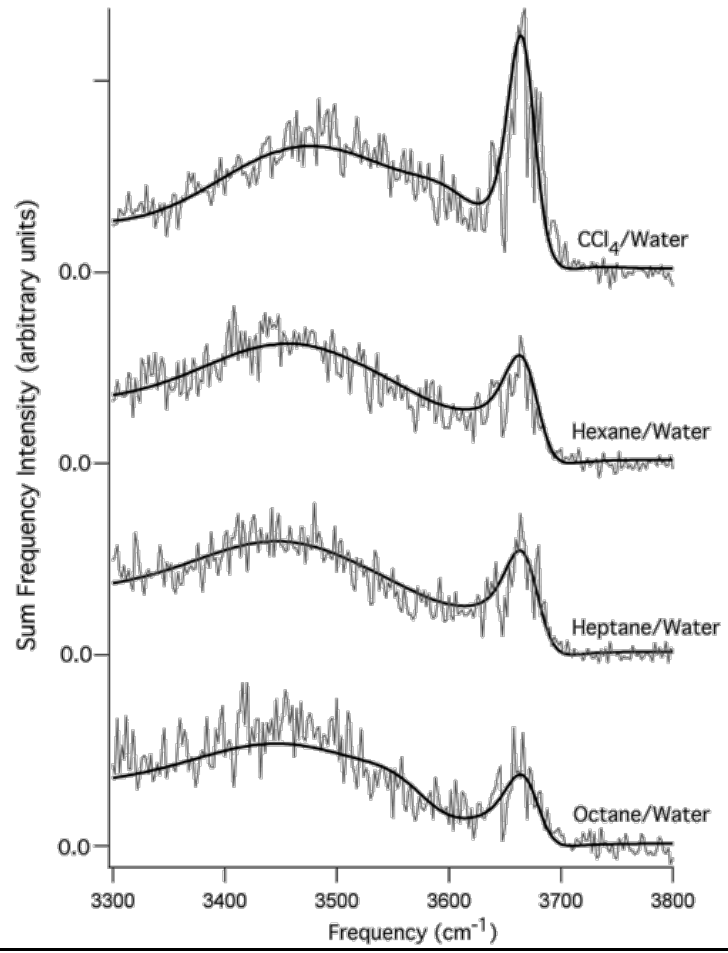 | *J. Phys. Chem. B* **2003,** *107* (1), 237-244 |
| 2014 | Bakker | Hexane/ $D_2O$ | Reflection SFG | Enhanced H-bonding in Hexane/$D_2O$ interface | 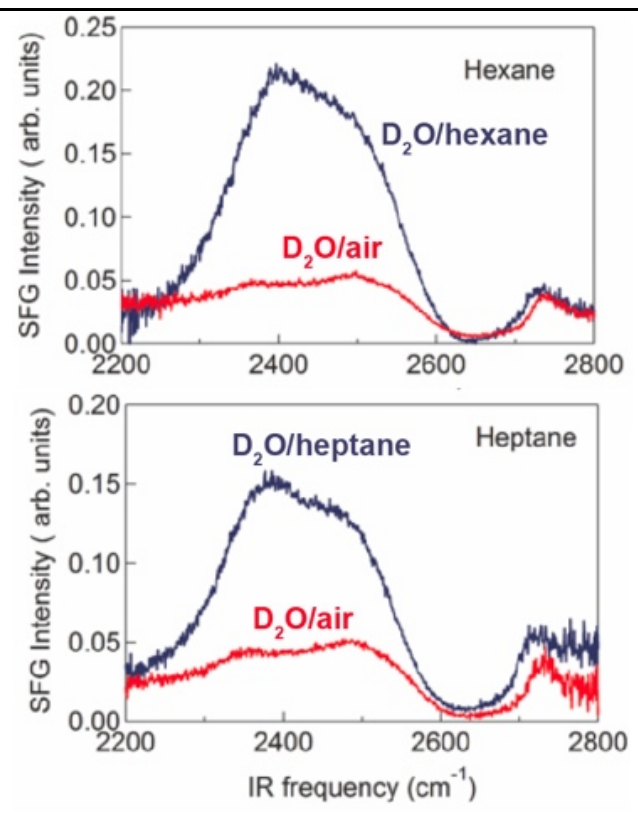 | *J. Chem. Phys.* **2014,** *140* (5), 054711 |
| 2020 | Tian | Hexane/ water | Reflection SFG | Stronger H-bonding | 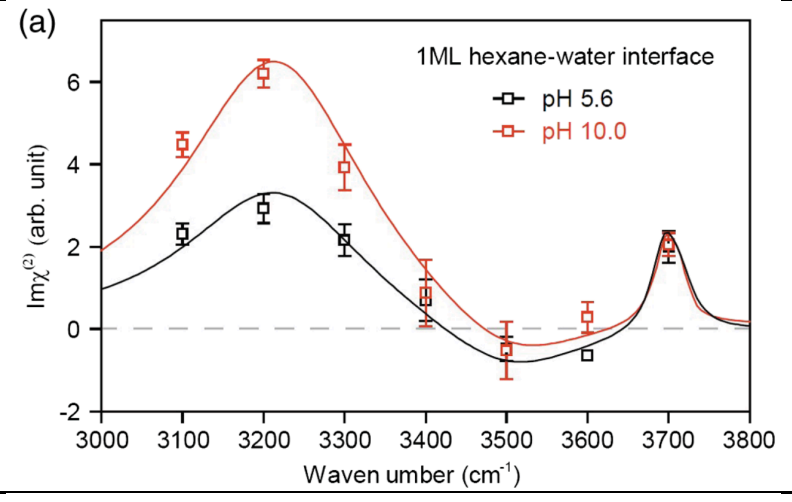 | *Phys. Rev. Lett.* **2020,** *125* (15), 156803 |
| 2020 | Roke | Oil droplets in $D_2O$ | SFS | Interfacial structure of water is more ordered; a high-frequency peak at 2735 $cm^{-1}$ is an artifact. | 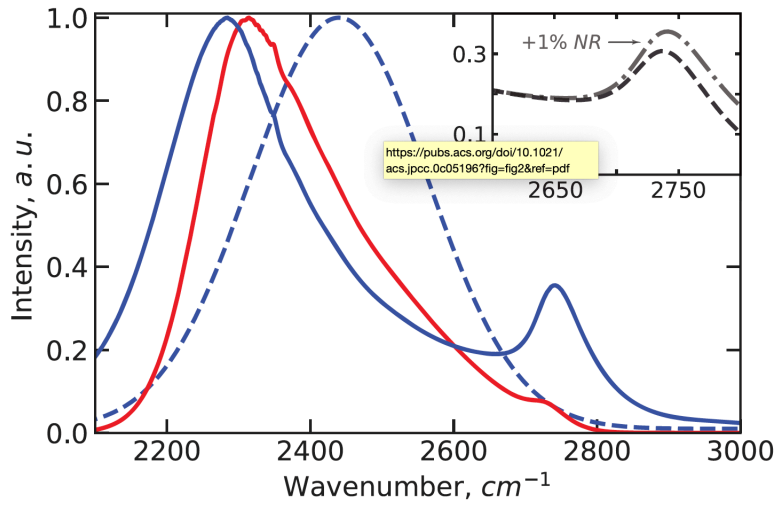 | *J. Phys. Chem. C* **2020,** *124* (42), 23078-23085 |
| 2021 | Richmond | Oil droplets in $D_2O$ | SFS | A peak just below 2700 $cm^{-1}$ (assigned to free OD) is unaffected after absorption correction. | 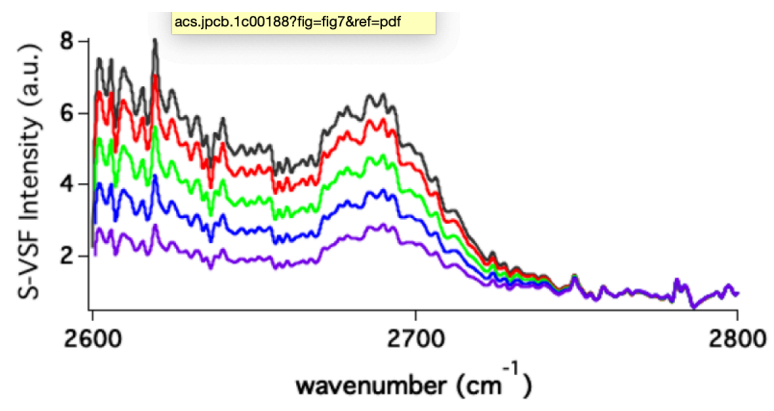 | *J. Phys. Chem. B* **2021,** *125* (12), 3216-3229 |
| 2021 | Roke | Oil droplets in $D_2O$ | SFS | Interfacial structure of water is more ordered; non-H-bonded OD mode is blue-shifted to ~2650 $cm^{-1}$ due to C-H···O hydrogen bonding. | 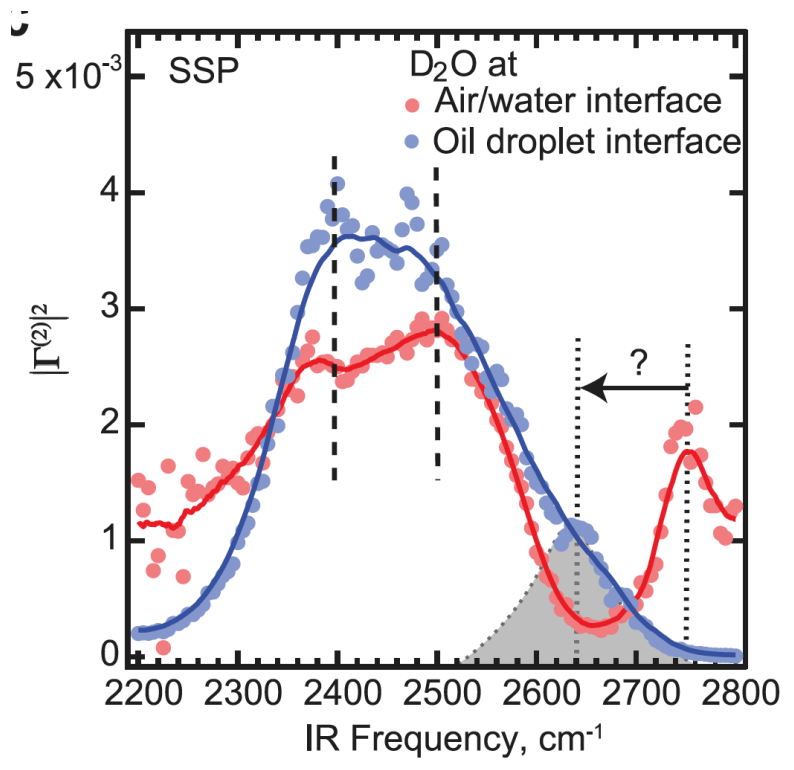 | *Science* **2021,** *374* (6573), 1366-1370 |

SFG: sum frequency generation

SFS: sum frequency scattering

**Supplementary Note 2. Discussion on the preparation of stable oil droplets**

We note there is a great deal of discussion on the origin of the large negative surface charge of bare (surfactant-free) oil emulsions[7-10]. Surface impurities are one of the hypotheses. Therefore, we also studied the effects of impurities and cleaning procedures on the properties of oil droplets. We thoroughly wash all glassware including glass vial, coverslip and glass slide with long-time acid incubation (like Piranha solution) followed by sufficient water sonication. We found with such thorough glass cleaning, the formation of milky emulsion is not easier but more difficult. The solution is totally transparent even after vortex and sonication. Though milky emulsion may still form after sufficient long sonication (like over an hour), the measured zeta potential (ZP) value is often smaller (~ -40 mV), the droplet size is larger (diameter>400 nm) and the sample is also more unstable (droplet coalescence within 1-2 days). Meanwhile, we observed a very quick formation of milky suspension using uncleaned vials. However, while the mean droplet size (diameter ~250 nm) and the PDI value (<0.2) are what we expected, the measured ZP value is still smaller (~ -40 mV).

Carpenter et al. stated the key to the creation of stable droplets is a rigorous glassware cleaning protocol[11]. In contrast, a more recent work from Pullanchery et al. showed that the cleaning procedure of glassware has no influence on the electrophoretic mobility of the droplet[12]. They proposed an alternative explanation that prolong acid treatment and subsequent ultrasonication will cause the detachment of glass nanoparticles from the glass surface in the emulsification process and alter the oil droplet surface chemistry[12]. Our own experimental experience agrees with their results that prolonged acid baths following by sonication may create emulsions with lower charge (i.e. smaller negative ZP value).

Inspired by the explanation from Pullanchery et al., we have tried rinsing the glassware with sufficient amount of Milli-Q water after Piranha wash. All glassware were further dried in a 90-degree oven to remove acid as much as possible. We also control the total sonication time in the emulsification process to within 15 minutes. In this way, we obtained reproducible preparation of oil emulsions with a consistent diameter (~250 nm) and a large negative ZP value (~ -60mV). Hence, a proper cleaning procedure is required to remove unknown impurities (likely some negative-charged species) in glass vial; in the meanwhile, over-cleaning may bring more troubles to the system.

We prepared the emulsion sample with 1% $C_{16}D_{34}$. In practical, we found the true concentration is at the level of 0.4-0.5%. Emulsion with much higher mole fraction becomes unstable. It is worth to mentioning that we usually observe the existence of a thin oil layer on top of the aqueous layer, indicating the true concentration of oil droplets in water is below the percentage we have used. This is also supported by the intensity of the measured Raman spectra in the C-H (or C-D) region. We also found the oil emulsion suspensions will cream out after a couple of days as the creamy droplets accumulate on the top of the solution. However, the solution can be redispersed by shaking, and the particle size and zeta potential value of the sample remains relatively stable.

**Supplementary Note 3. A literature overview on the dangling OH peak in different systems.**

First, dangling OH peak is sensitive to the local environment of the OH bond including the charge. For instance, Raman-MCR revealed narrow dangling peak with a frequency of ~3660 $cm^{-1}$ in the hydration shell of alcohols[13]. In contrast, this peak could be red-shifted to ~3641 $cm^{-1}$ for positively charged tetraphenylarsonium ($TPA^+$) ion[14], ~3610 $cm^{-1}$ for benzene due to pi-hydrogen bonding[1], and ~3577 $cm^{-1}$ for negatively charged tetraphenylborate ($TPB^-$) ion[14], respectively. Second, without charge, most dangling OH peaks seem to converge at ~3650 $cm^{-1}$ which is close to the water vapor Raman peak. More information about the position of the dangling peak in different systems is summarized below.

| System | Technique | Peak center (cm-1) | Spectra | Ref. |
|---|---|---|---|---|
| macroscopic vapor-water interface | SFG | 3705 | 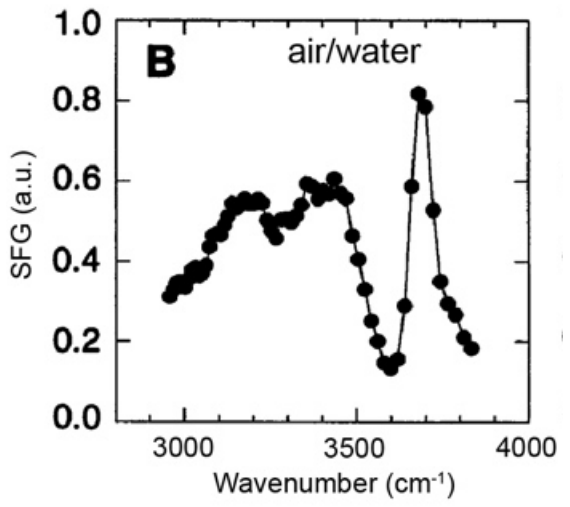 | *Science* **1994,** *264* (5160), 826-828 |
| macroscopic oil-water interface | | 3674 | 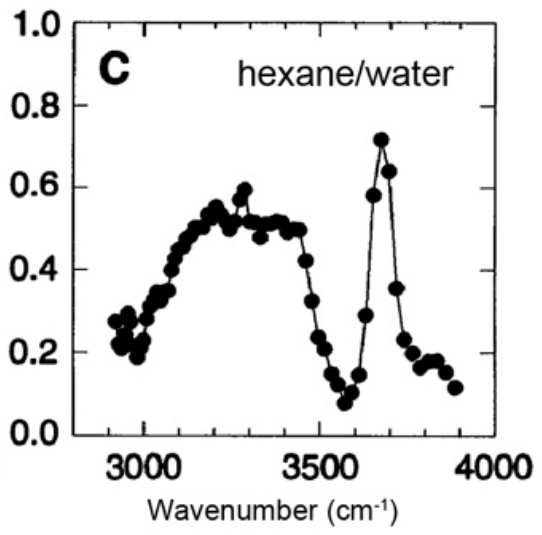 | |
| Hydration shell of alcohols | Raman-MCR | ~3661 | 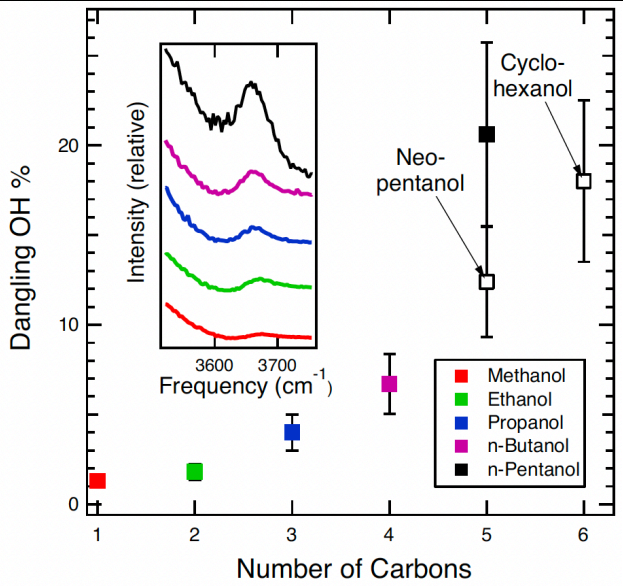 | *Proc. Nat. Acad. Sci. U.S.A.* **106**, 12230-12234 (2009) |
| Hydration shell of $CO_2$ | Raman-MCR | 3654 | 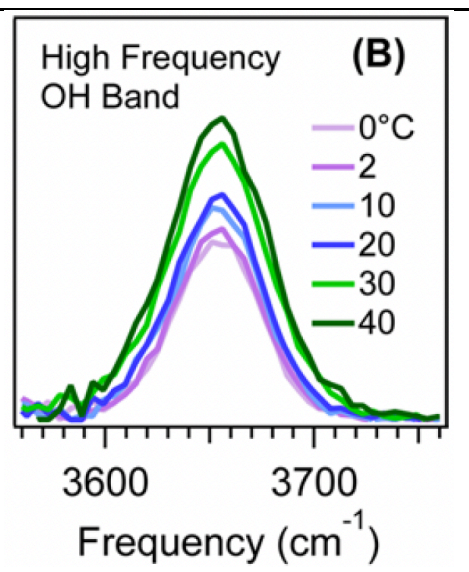 | *J. Phys. Chem. Lett. 2017, 8, 2971−2975* |

<table>
<tr><td>Hydration shell of benzene</td><td>Raman-MCR</td><td>3600</td><td>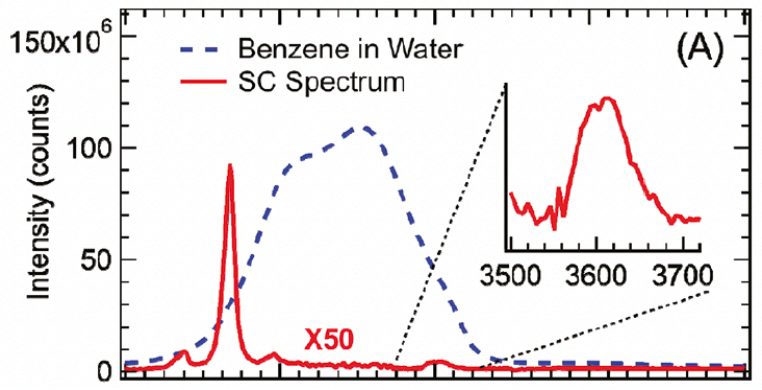
</td><td>J. Phys. Chem. Lett. 2, 2930-2933 (2011)</td></tr>
<tr><td>Hydration shell of tetraphenylarsonium (TPA$^+$) ion</td><td rowspan="2">Raman-MCR</td><td>3641</td><td rowspan="2">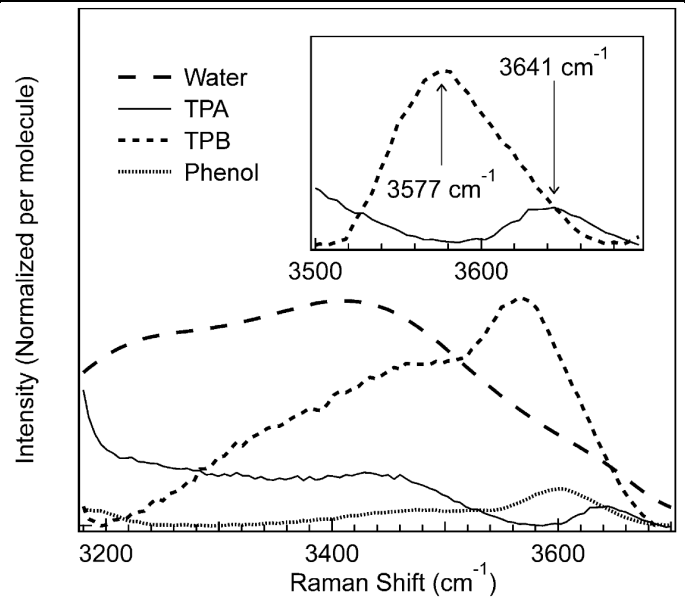
</td><td rowspan="2">Angew. Chem. Int. Ed. 53, 9560-9563 (2014)</td></tr>
<tr><td>Hydration shell of tetraphenylborate (TPB$^-$) ion</td><td>3577</td></tr>
<tr><td>Liquid water (290 K, 0.1MPa)</td><td>Spectral fitting of Raman spectroscopy</td><td>3636</td><td>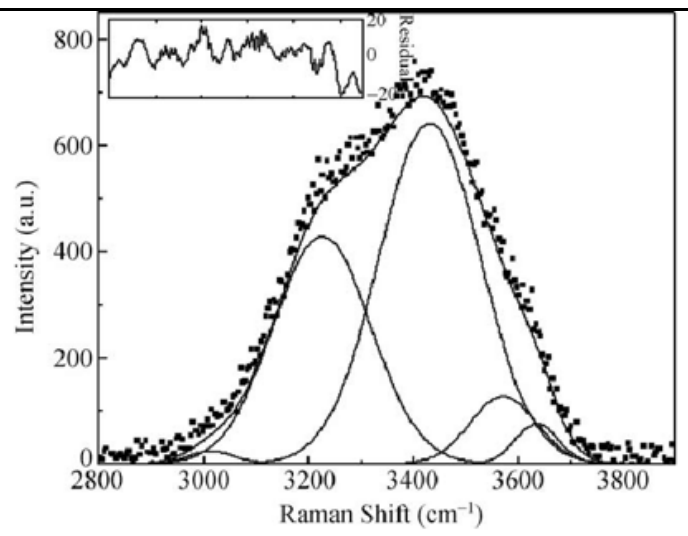
</td><td>Vibrational Spectroscopy 51, 213-217 (2009).</td></tr>
<tr><td>Water vapor</td><td>Raman spectroscopy</td><td>3657</td><td>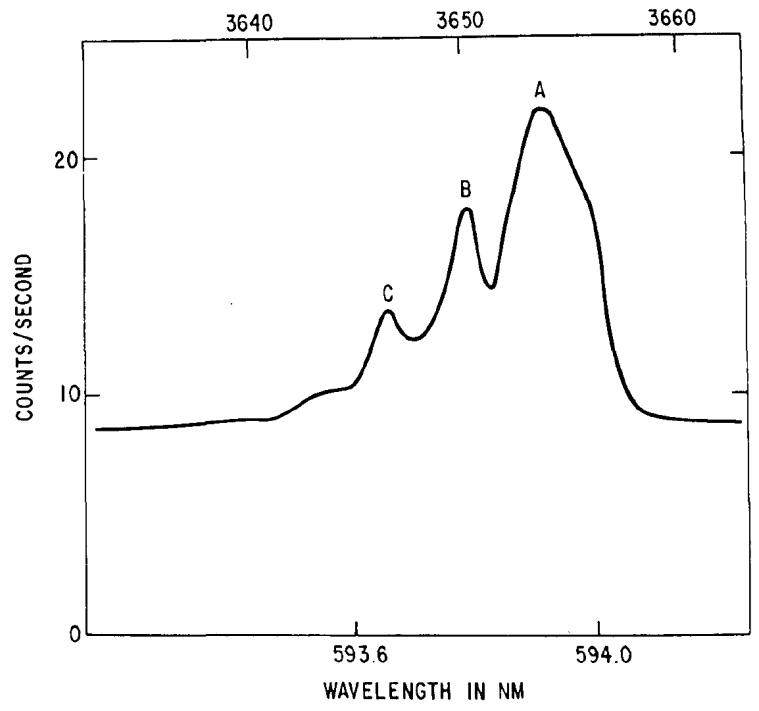
</td><td>J. Opt. Soc. Am. 66, 422-425 (1976).</td></tr>
</table>

## SUPPLEMENTARY FIGURES

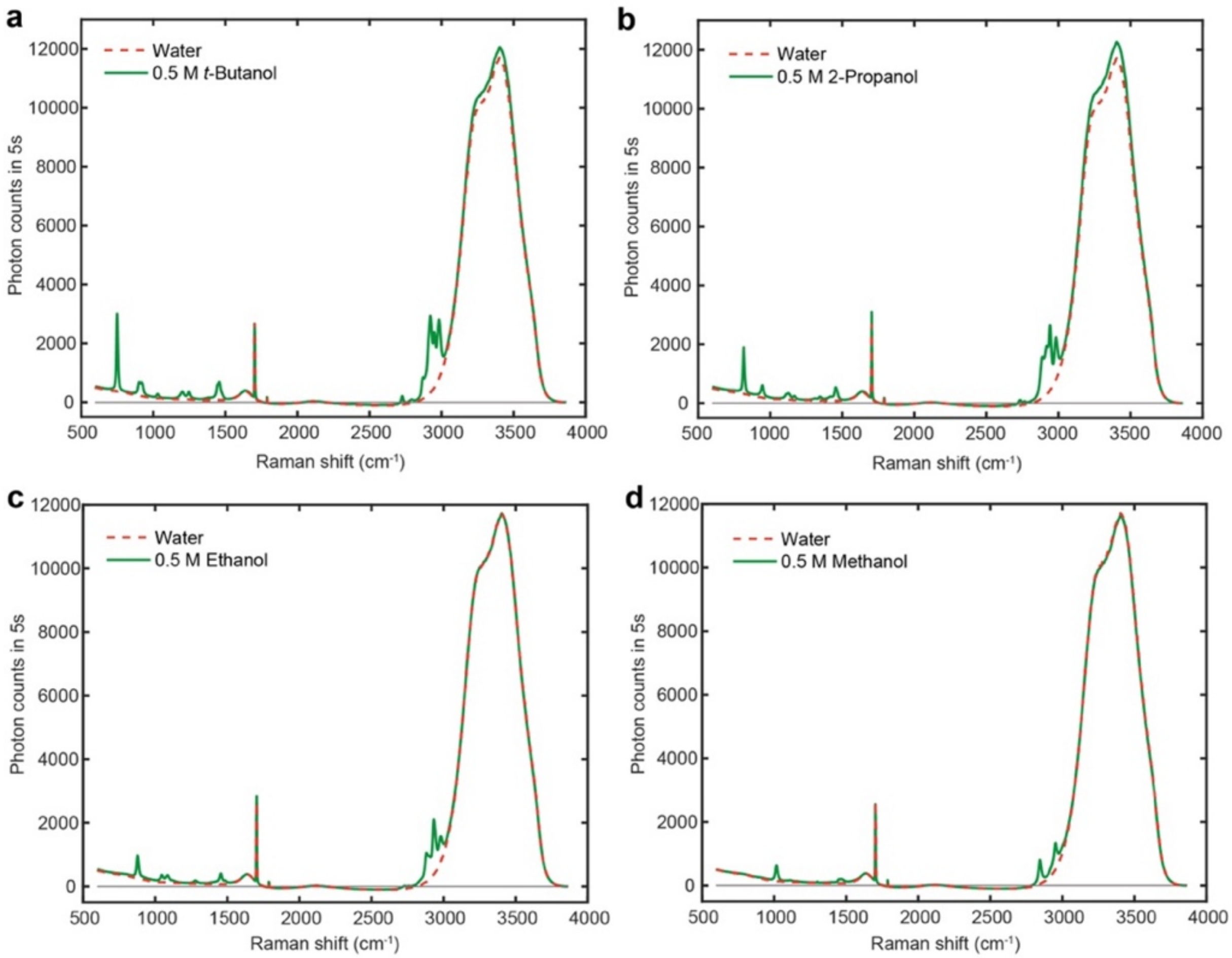

**Supplementary Figure 1. Raman spectra of pure water and aqueous alcohol solutions.** (a) 0.5 M *t*-butanol, (b) 0.5 M 2-propanol, (c) 0.5 M ethanol, (d) 0.5 M methanol.

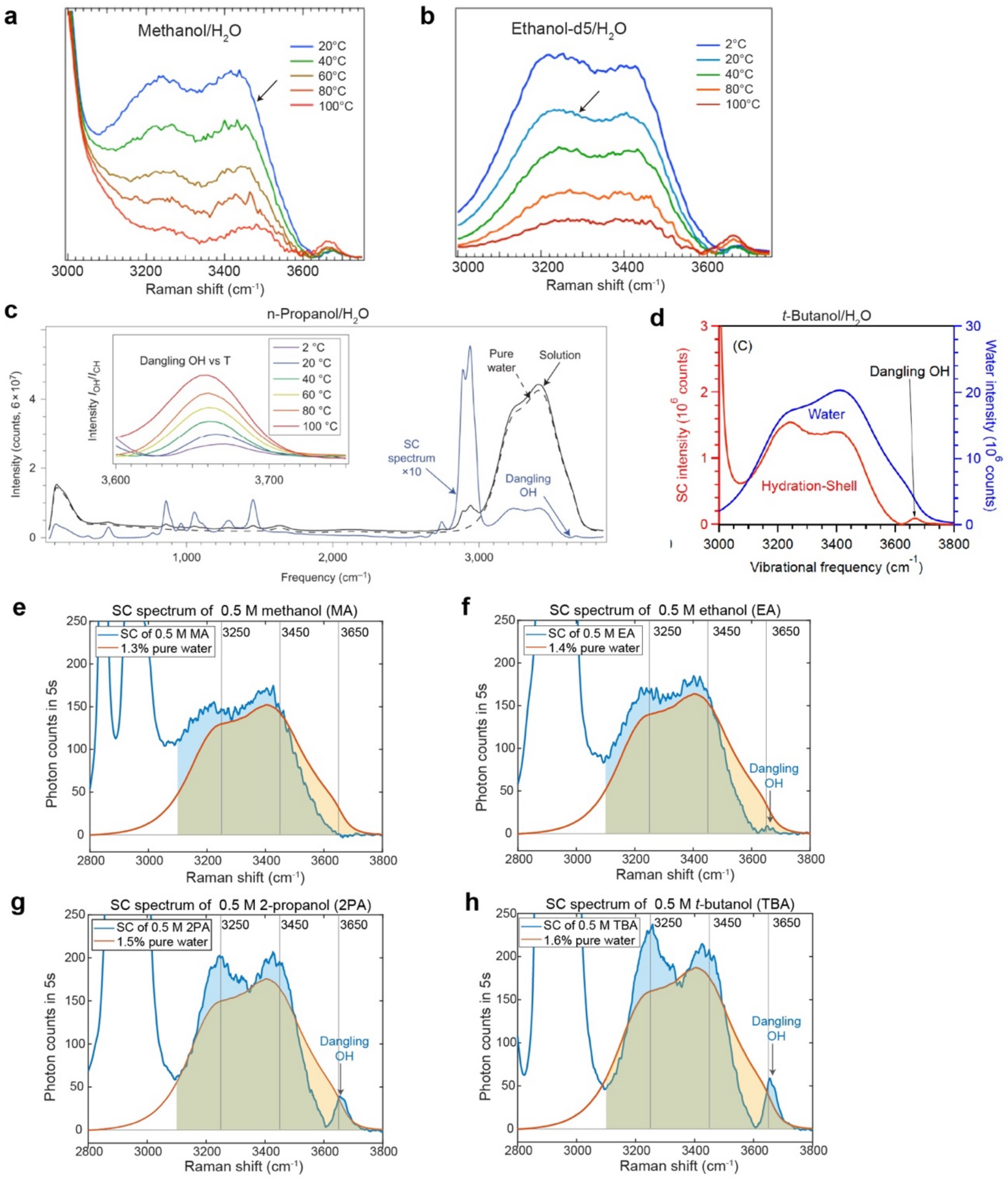

**Supplementary Figure 2. Solute-correlated (SC) spectra of aqueous alcohol solutions in literature and from our instrument.** (a) Raman-MCR SC spectra (solid curves) obtained from ~0.5 M aqueous methanol solutions at temperatures between 20ºC and 100ºC. (b) Raman-MCR SC spectra (solid curves) obtained from ~0.5 M aqueous solutions of deuterated ethanol at temperatures between 2ºC and 100ºC. (c) Raman spectra of pure water (dashed black curve) and 0.5 M aqueous 1-propanol (solid black curve) at 20 °C, as well as the resulting Raman-MCR SC spectrum (blue curve) of 1-propanol. The inset shows an expanded view of the dangling OH bands obtained at various temperatures, each normalized by dividing the SC spectrum by the corresponding CH band area. (d) Raman-MCR decomposition for 1 M tert-butyl alcohol (TBA) at 20 ºC. (e-h) SC spectra measured from our instrument of (e) 0.5 M methanol, (f) 0.5 M ethanol, (g) 0.5 M 2-propanol, (f) 0.5 M *t*-butanol. See Supplementary Fig. 1 for the raw data. (a-c) are adopted from *Nature* **491**, 582-585 (2012); (d) is adopted from *Proc. Nat. Acad. Sci. U.S.A.* **106**, 12230-12234 (2009).

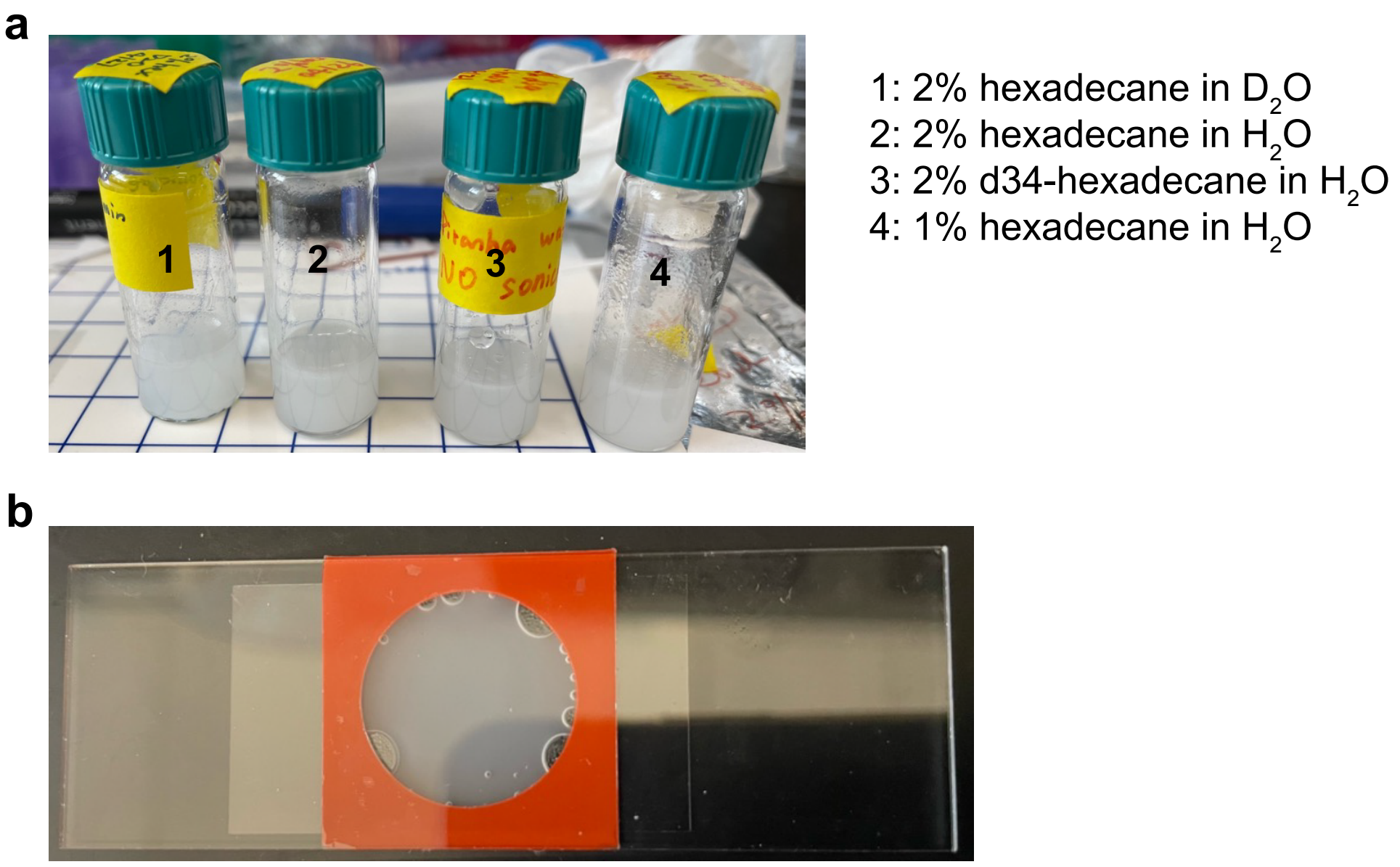

**Supplementary Figure 3. Examples of oil droplets emulsion sample. (**a) Oil droplets prepared with different concentrations. (b) A sandwiched oil emulsion sample for confocal Raman micro-spectroscopy measurement.

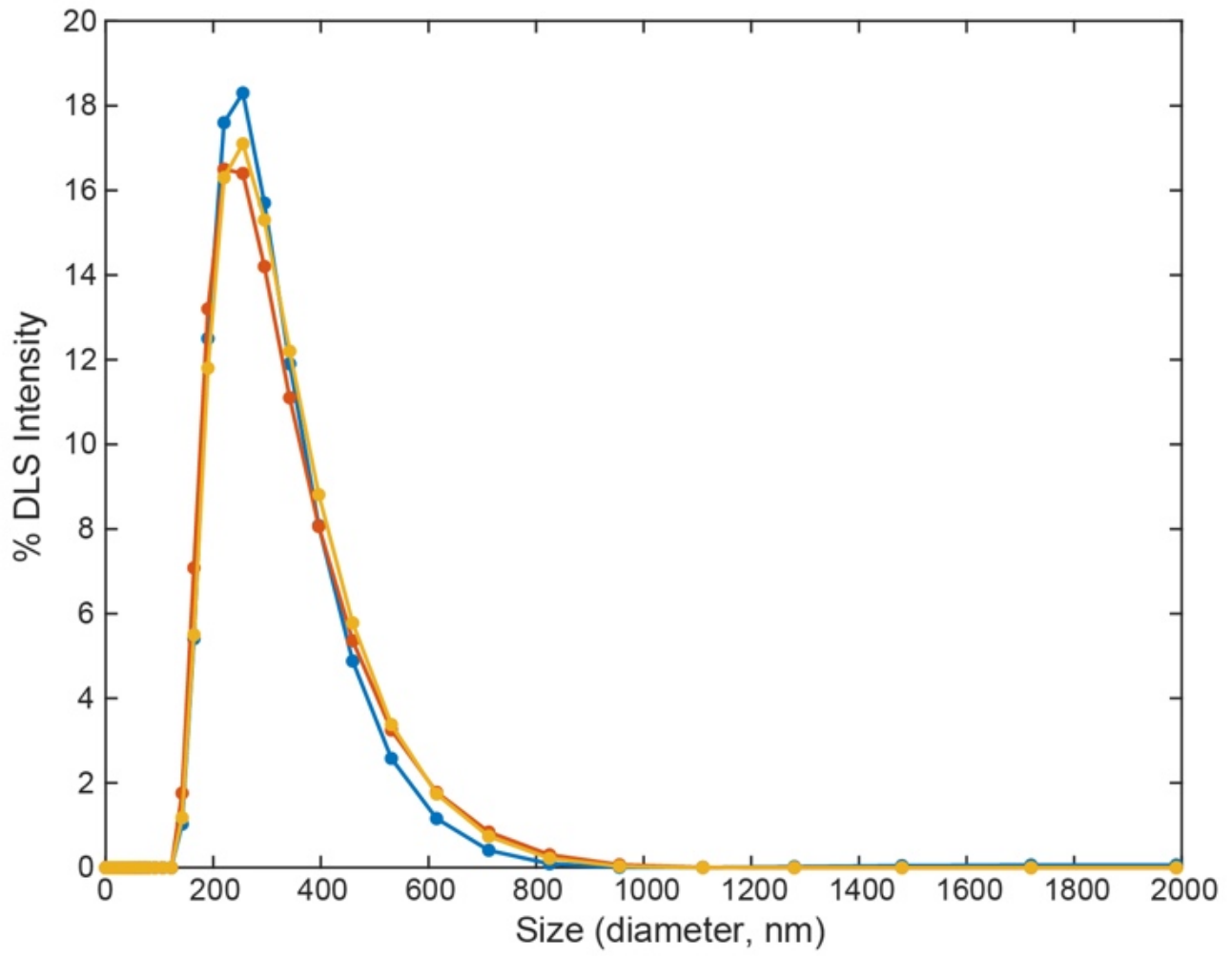

**Supplementary Figure 4. Size distribution of oil emulsions.** DLS intensity-diameter distribution of 2% hexadecane droplets in ultrapure water.

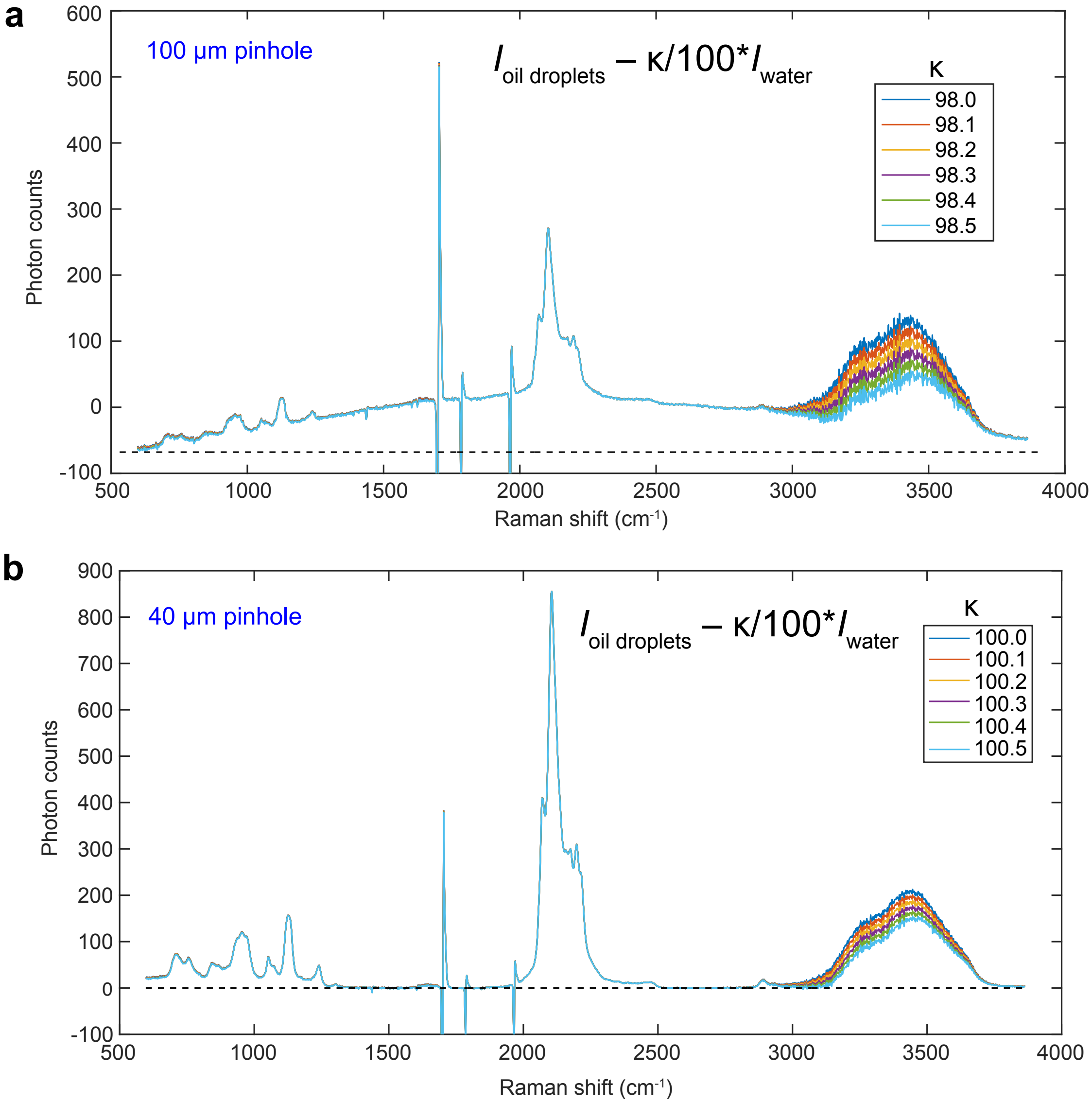

**Supplementary Figure 5**. Comparison of the Raman difference spectra with different pinhole sizes. Difference Raman spectra of 2% d34-hexadecane oil droplets using (a) a larger pinhole size of 100 μm, evident scattering background as an uneven baseline after subtraction is observed and (b) In contrast, the baseline after subtraction under a 40 μm pinhole is much flatter on a same sample. Various ratio values ($\kappa$) were used to calculate the difference spectra between oil droplets and pure water.

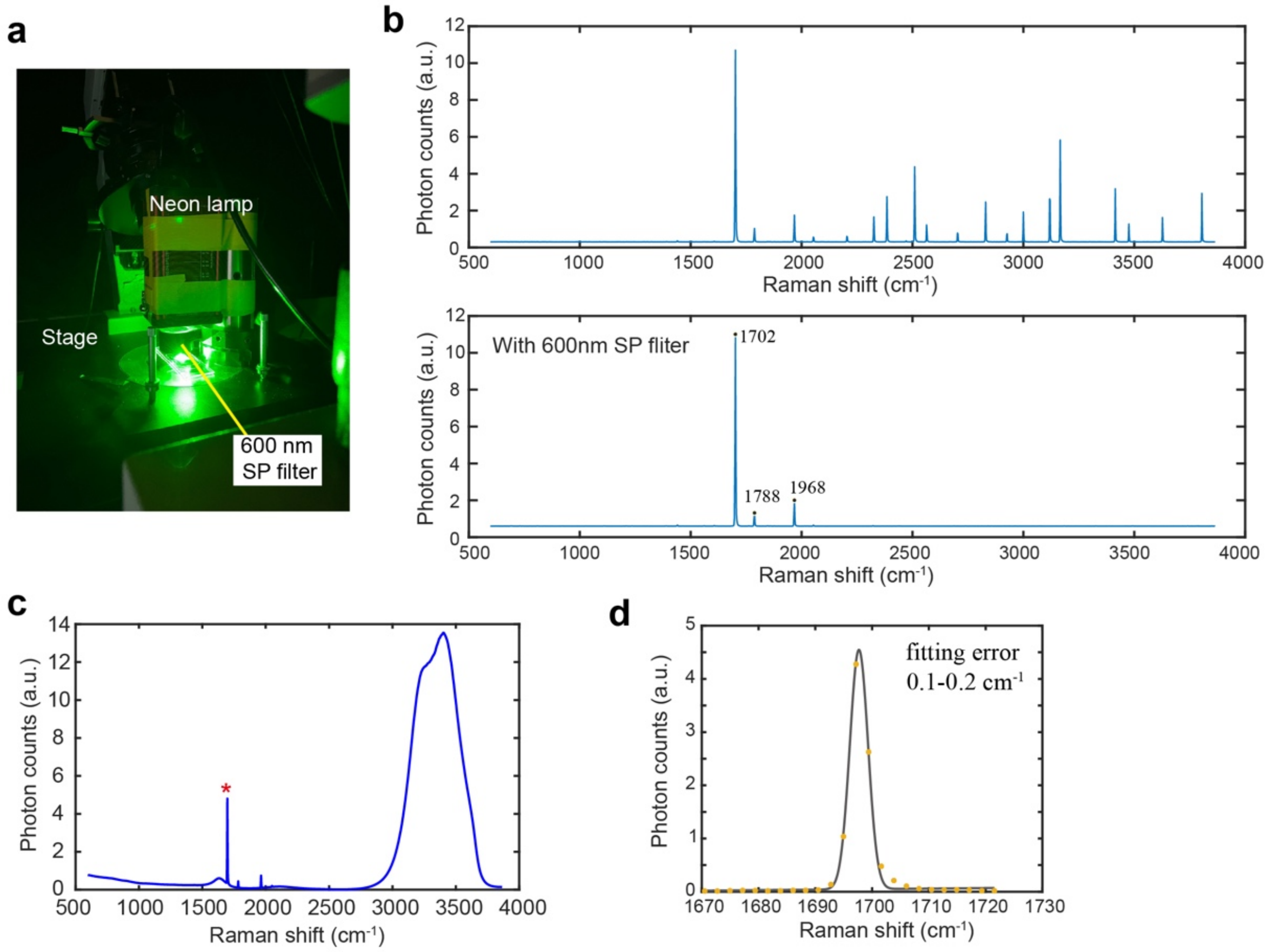

**Supplementary Figure 6**. **Sub-pixel spectral correction with narrow emission lines from a Neon lamp.** (a) setup of Neon lamp on home-built Raman system for sub-pixel spectral correction. A Neon lamp is placed above the sample stage, and the transmitted light is measured with the detection system of confocal Raman. A 600-nm short-pass filter is added to block emission peaks after 2100 $cm^{-1}$. (b) (up) typical neon emission spectrum and (down) neon emission spectrum with a 600-nm short-pass filter. (c) Raman spectrum of water. Asterisk indicates the used emission line (~1700 $cm^{-1}$) for Raman shift correction. (d) Fitting the ~1702 $cm^{-1}$ emission line with a Gaussian profile for spectral correction.

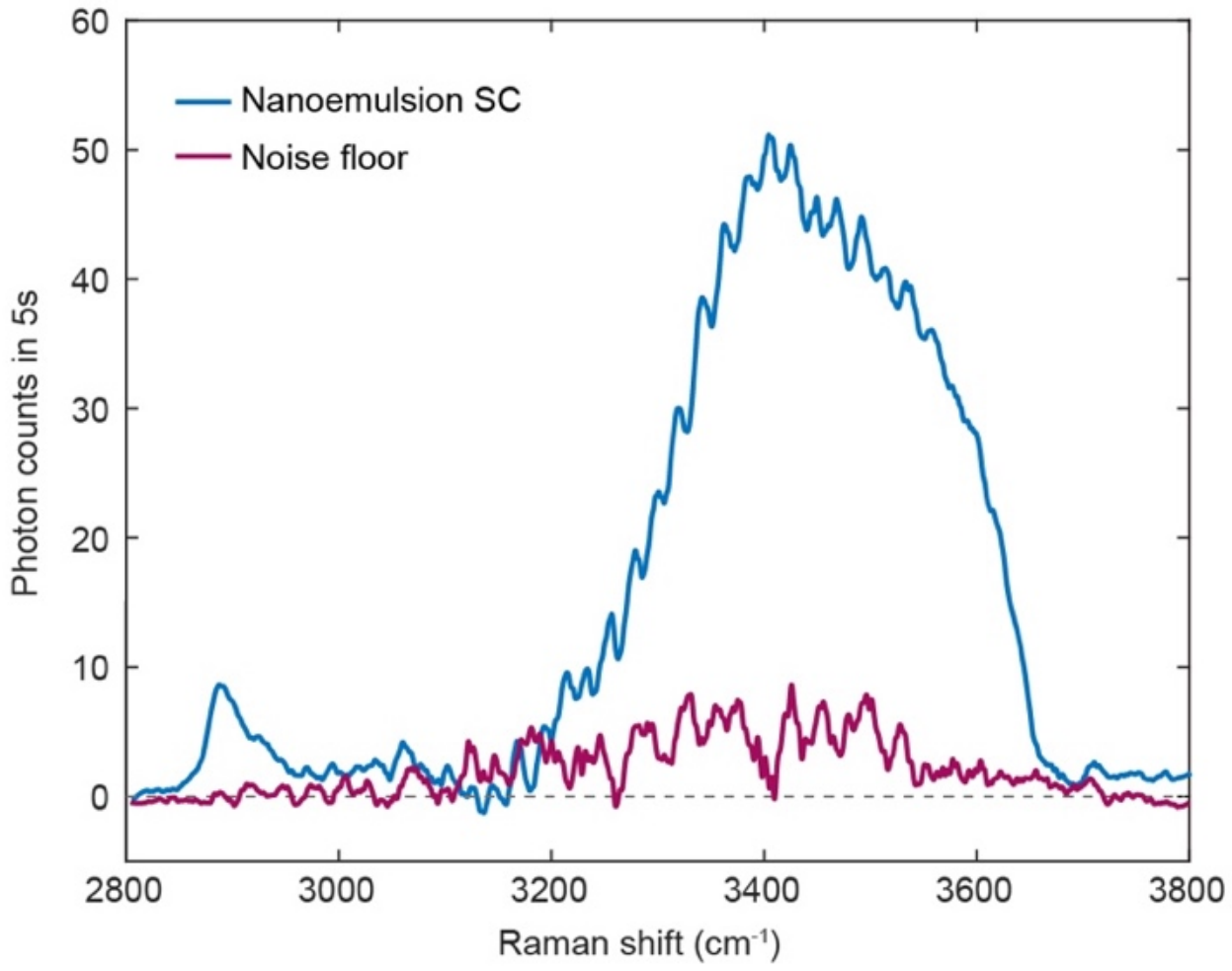

**Supplementary Figure 7. Noise level after sub-pixel spectral drift correction.** Blue line: the SC spectrum of 1% d34-hexadecane oil emulsion. Purple line: SC spectrum from a 'water-subtract-water' negative control. We performed two parallel measurements on the pure water sample, treated one measurement as the solvent spectrum and the other as solution spectrum. After sub-pixel spectral drift correction, the noise level of the negative control is approaching the technical noise level and is way lower than our measured SC spectrum of oil emulsion. This comparison supports the experimental reliability.

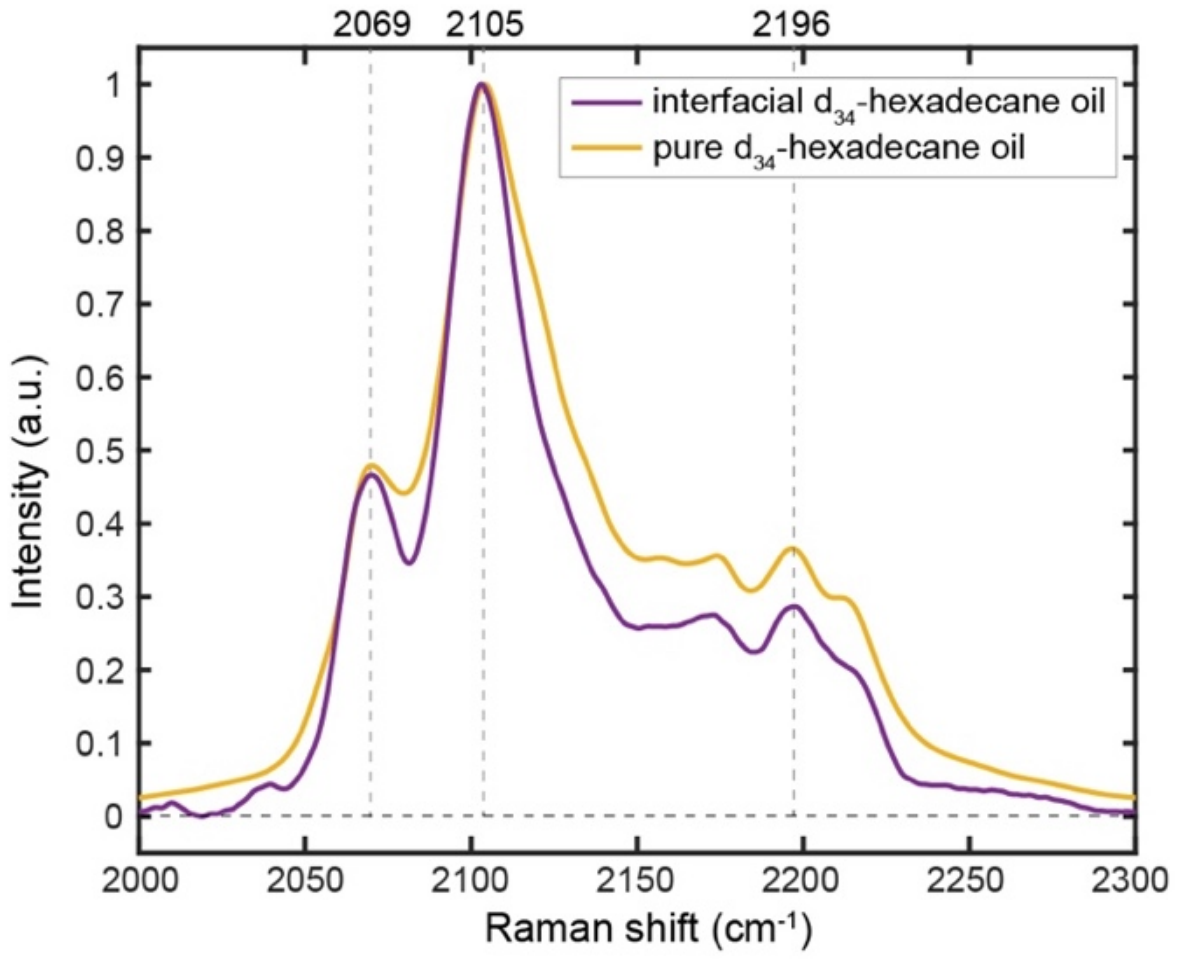

**Supplementary Figure 8. Water-perturbed, solute-correlated oil spectrum of d34-hexadecane oil droplets.** While Raman-MCR spectroscopy has mostly focused on the water spectrum, the strategy should be general and thus extendable to the oil phase. We therefore measured "water-perturbed" interfacial oil spectrum with MCR (purple), and compared it to a pure d34-hexadecane (yellow). Although there is a narrowing of multiple peaks at the oil-water interface, the positions of the peaks are largely consistent. The largest difference is that the "water-perturbed" oil spectrum appears more narrow, and the peak ratio of 2105 $cm^{-1}$ over 2196 $cm^{-1}$ is increased compared to bare oil, both suggesting the interfacial oil may be more ordered. Indeed, similar behavior is commonly observed in lipids at low temperature and more ordered phase[15,16]. An increase in molecular order can also be understood by the fact that oil molecules tend to lie parallel to the interface with water, which has been reported elsewhere for hexadecane-water and other oil-water systems[17-19]. More notably, we find that the water interface does not shift the 2105 $cm^{-1}$ symmetric stretch peak (nor any other peak) of the oil, contradicting measurements using VSFS by Pullanchery et al. who observed a ~5-10 $cm^{-1}$ shift[20].

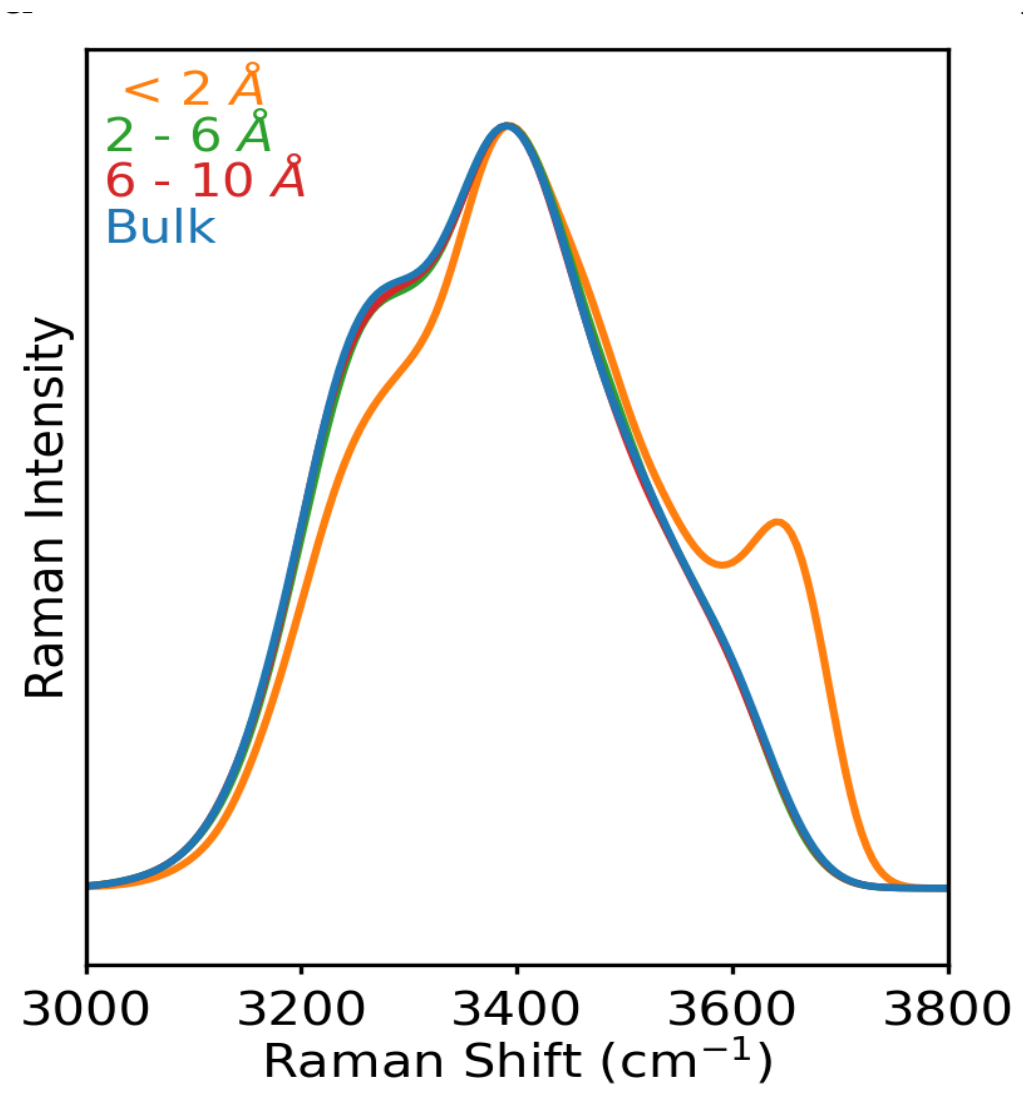

**Supplementary Figure 9. Raman spectra produced by theory as a function of distance from the WCI.** We show Raman spectra produced by the monomer-field model using only water within 2 Å of the WCI, within 2 – 6 Å of the WCI, and within 6 – 10 Å of the WCI. We also show the bulk spectra for reference. The interfacial spectra have converged to bulk-like within 2 Å of the WCI.

## Supplementary References